\documentclass{jfm}
\usepackage{bm,amsmath,amssymb,natbib,graphicx,mathrsfs,nicefrac}
\usepackage{array,epic,eepic,color}

\graphicspath{{./final_figures/}}

\newlength{\piclen}

\title[Boundary-layer turbulence  
in experiments of quasi-Keplerian flows]
{Boundary-layer turbulence  
in experiments of quasi-Keplerian flows}

\author[Jose M. Lopez and Marc Avila] {J\ls O\ls
  S\ls E\ns M.\ns L\ls O\ls P\ls E\ls Z,$^{1,2,3}$  and
  M\ls A\ls R\ls C\ns A\ls V\ls I\ls L\ls A$^{1,4}$}

\affiliation{ $^1$Institute of Fluid Mechanics,
  Friedrich-Alexander-Universit\"at Erlangen-N\"urnberg, 91058
  Erlangen, Germany\\[\affilskip] $^2$Departament de F{\'\i}sica Aplicada,
  Univ. Polit\`ecnica de Catalunya, Barcelona 08034,
  Spain\\[\affilskip] $^3$ Institute of Science and Technology,
  3400 Klosterneuburg, Austria\\[\affilskip] $^4$ Center of Applied Space Technology and Microgravity,
  University of Bremen, 28359 Bremen, Germany\\[\affilskip]}

\pubyear{} \volume{} \pagerange{}
\date{\today}
\begin{document}

\maketitle

\begin{abstract} 

Most flows in nature and engineering are turbulent because of their large velocities and spatial scales.
Laboratory experiments of rotating quasi-Keplerian flows, for which the angular velocity decreases radially but the angular momentum increases, 
are however laminar at Reynolds numbers exceeding one million. 
This is in apparent contradiction to direct numerical simulations showing that in these experiments turbulence transition is triggered by the axial boundaries. 
We here show numerically that as the Reynolds number increases turbulence becomes progressively confined  
to the boundary layers and the flow in the bulk fully relaminarizes.
Our findings support that turbulence is unlikely to occur in isothermal constant density quasi-Keplerian flows.

\end{abstract}

\section{Introduction}

\noindent Understanding the origin of turbulence in accretion disks is a long-standing problem in astrophysics~\citep{JiBa13}.
The simplest model for the flow of gas in an accretion disk consists of an isothermal incompressible constant density fluid
rotating with a Keplerian angular velocity $\Omega \propto r^{-3/2}$, where $r$ is the distance to the accreting central object.
Despite the hydrodynamic stability of such flows to small disturbances~\citep{Ra17},
the possibility of a nonlinear transition to turbulence via finite-amplitude disturbances is not precluded.
However, this has not been demonstrated and so several mechanisms capable of destabilizing Keplerian flows have been proposed
in the literature \citep[see][for a recent review]{Tuetal14}.
Prominent amongst these is the magnetorotational instability~\citep[MRI, see][]{BaHa98,Ba03},
which can drive vigorous turbulence and transport angular momentum  at rates required for accretion to occur.
However, the MRI operates in ionized disks only and so it does not apply to cool protostellar disks.
There are many mechanisms that might give rise to turbulence in the absence of magnetic fields:
baroclinic instabilities~\citep{KlaBo03,JoGa06,PeStJu07},
instabilities driven by radial~\citep{GoSch67,Fr68,UrBra98} or axial stratification~\citep{ShaRu05,DuMaNoRiHeZa05,MaPeJiBaLe15},
crossflow instabilities~\citep{Ker15}, convective instabilities~\citep{LiPa80,RyGo92} or self-gravitation~\citep{To64,LiPri87}.
Nevertheless, their applicability to accretion disks is still under investigation.
Whereas there is an understanding of the underlying instability mechanisms,
their non-linear evolution and saturated state have still to be studied in realistic disk simulations
with the proper radiation transport and analysing the effect of the boundary conditions.

\noindent In general, the lack of observational evidence and the computational limitations in simulating the extreme
parameter values governing the dynamics of accretion disks makes them a particularly difficult object to study \citep[see e.g.][]{Miesch15}.
This has motivated the development of laboratory experiments capturing the essential physics at play.
Quasi-Keplerian flows, for which the angular velocity decreases radially, whereas the angular momentum increases, can in principle be realized
in laboratory experiments of fluids between two concentric rotating cylinders (Taylor--Couette flow).
If the cylinders are assumed to be infinite in length the basic laminar flow is purely azimuthal 
\begin{equation}\label{Couetteflow}
  v(r)=\frac{\Omega_o r_o^2-\Omega_ir_i^2}{r_o^2-r_i^2}r+
  \frac{(\Omega_i-\Omega_o)(r_ir_o)^2}{r_o^2-r_i^2}\frac{1}{r},
\end{equation}
\noindent where $\Omega_i$ ($\Omega_o$) and $r_i$ ($r_o$) are the angular velocity and radius of the inner (outer) cylinder.
Provided that $\Omega_i>\Omega_o$ but $r_i^2\Omega_i<r_o^2\Omega_o$ then the basic \emph{Couette} flow~\eqref{Couetteflow} is quasi-Keplerian.
Despite the apparent simplicity of this model,  laboratory realizations of quasi-Keplerian velocity profiles are fraught with difficulty.
In practice, the viscous interaction between fluid and end-plates confining the fluid in the axial direction results in secondary flows,
also known as Ekman circulation (EC). EC can extend deep into the bulk flow and cause the azimuthal velocity to significantly
deviate from the theoretical profile~\eqref{Couetteflow} as the rotation speeds increase~\citep{RiZa99,hollerbach2004}.

\noindent Pioneering experiments of quasi-Keplerian flows were conducted by \cite{JiBuSchGo06}, who used a short height-to-gap aspect ratio and end plates
split into two independently rotating rings. \cite{JiBuSchGo06}  carefully adjusted the rotation speed of the rings so as to minimize EC and
measured Reynolds stresses in the bulk. They concluded that the bulk flow was laminar despite reaching Reynolds numbers of up to $10^6$.
These results were questioned by \cite{PaLa11}, who used  a tall apparatus with end plates attached to the outer cylinder.
\cite{PaLa11} split the inner cylinder vertically in three sections and measured the torque on the central section so as to reduce
the effect of the end plates in their measurements. However, \citet{Av12} performed direct numerical simulations
reproducing the precise geometry of  \cite{JiBuSchGo06} and \cite{PaLa11} and showed that in both setups strong EC render the
flows turbulent at Reynolds numbers as low as $10^3$. Although his simulations were consistent with the turbulent flows
observed by \cite{PaLa11}, as confirmed by subsequent experiments~\citep{NoHuVdVSunLoLa15}, they were in apparent
contradiction with the laminar flows observed by \cite{JiBuSchGo06}. 

 More recently, \cite{EdJi14} attributed this discrepancy to the large gap in Reynolds numbers between simulations and experiments.
They directly measured velocity profiles in a new experimental setup and showed that if the end plates are rotated within a certain range of velocities,
hereafter referred to as optimal boundary conditions, quasi-Keplerian Couette flow \eqref{Couetteflow} is obtained and remains stable even
when subject to strong disturbances.  

 We here perform direct numerical simulations of experimental quasi-Keplerian flows for Reynolds numbers up to $50,000$.
We show that as the Reynolds number increases, turbulence becomes progressively confined to thin boundary layers at the end plates.
As a result laminar quasi-Keplerian profiles are realized  in the bulk of the experiment. We provide a detailed picture of the relaminarisation
process and the role of the boundary conditions for two distinct configurations studied by \cite{EdJi15}.
Our results bridge the gap between experiments and previous simulations and
support the experimental conclusion that constant density isothermal quasi-Keplerian Taylor--Couette flows are  stable.

\section{Specification of the problem and numerical methods}                                                                                                                       
                                                                                                     
 A fluid of kinematic viscosity $\nu$ is contained in                                      
the annular gap between two vertical                                                      
concentric cylinders of length $h$ and radii $r_i$ and $r_o$.                             
The subindex $i$ ($o$) denotes the inner (outer) cylinder and $d = r_o-r_i$ is the gap width.                                
Differential rotation is generated by rotating the cylinders                                          
at independent angular velocities $\Omega_i$ and $\Omega_o$.     
The shear Reynolds number, $R_s$, and the rotation number, $R_\Omega$,
were chosen as control parameters
\begin{equation}\label{control_params}
  R_s = \dfrac{\tilde{S} d^2}{\nu}, 
  \qquad
  R_\Omega = \dfrac{2\tilde{\Omega}}{\tilde{S}},
  \end{equation}
where $\tilde{S}$ and $\tilde{\Omega}$ are the shear and rotation speed of the basic Couette flow \eqref{Couetteflow}
evaluated at the mean geometric radius $\tilde{r}=\sqrt{r_i r_o}$. The rotation number allows for a clear distinction between
cyclonic ($R_\Omega > 0$) and anticyclonic  ($R_\Omega < 0$) flows.
It has been widely used to characterize different rotation regimes of the Taylor--Couette                    
experiments and to compare results for different geometries~\citep{DuDaDaLoRiZa05,RaDeWe10,PaVgDuSunLo12}.     

In astrophysics  $q = -\frac{d \ln \Omega}{d \ln r}$ is used to characterise the rotation law.
The flow is quasi-Keplerian if $R_\Omega$ ($q$) is chosen within the range                
$-\infty < R_\Omega < -1$ ($0 < q < 2$). In the simulations presented in this paper,      
$R_\Omega = -1.038$ ($q = 1.93$) was chosen, corresponding to the experiments of~\cite{JiBuSchGo06}
and simulations of~\cite{Av12}, and close to the value $q = 1.8$  chosen by \cite{EdJi14}.
Their apparatus is axially bounded by two horizontal plates
which rotate differentially with respect to the cylinders.
These end plates can be further split into
several independently rotating rings whose angular velocities
can be adjusted to best approximate~\eqref{Couetteflow}.
The geometry of their apparatus is fully specified by                                       
two dimensionless parameters: the radius ratio,                                           
$\eta = r_i/r_o = 0.3478$, and the length-to-gap aspect ratio,                            
$\Gamma = h/d = 2.1$.                                                                                                                                           

\begin{figure}\setlength{\piclen}{0.35\linewidth}                                                      
  \begin{center}                                                                                       
    \begin{tabular}{ccc}                                                                               
      $(a)$  &  $(b)$ & $(c)$  \\                                                                      
      \includegraphics[width=0.65\piclen]{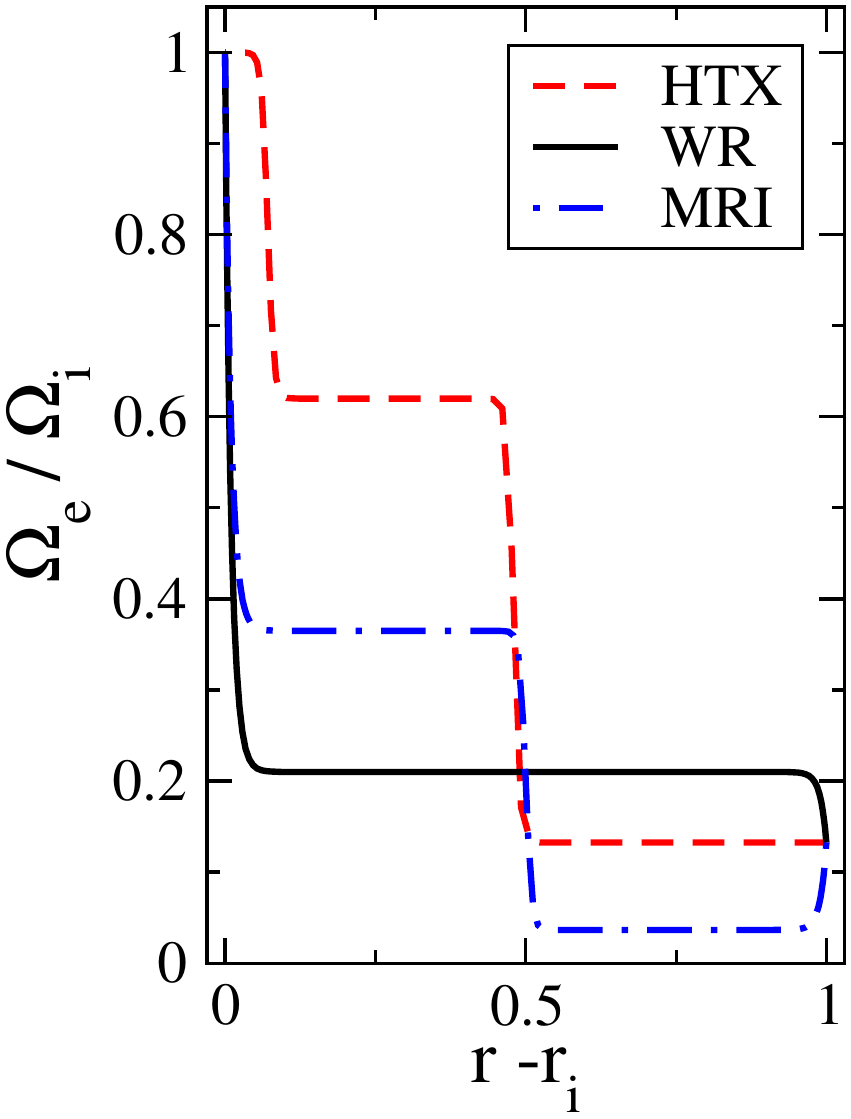} &                                                      
      \includegraphics[width=\piclen]{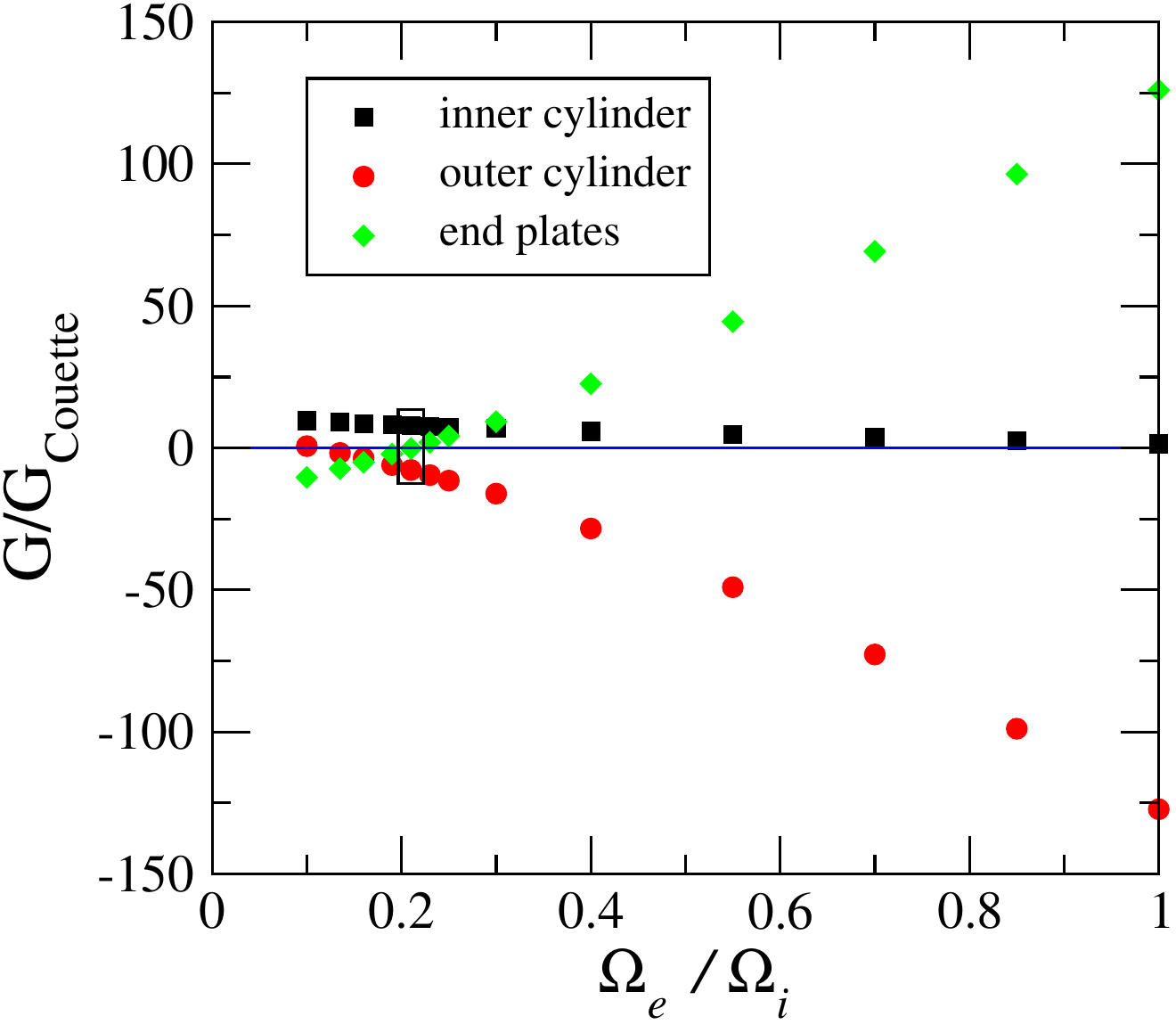} &                                                   
      \includegraphics[width=\piclen]{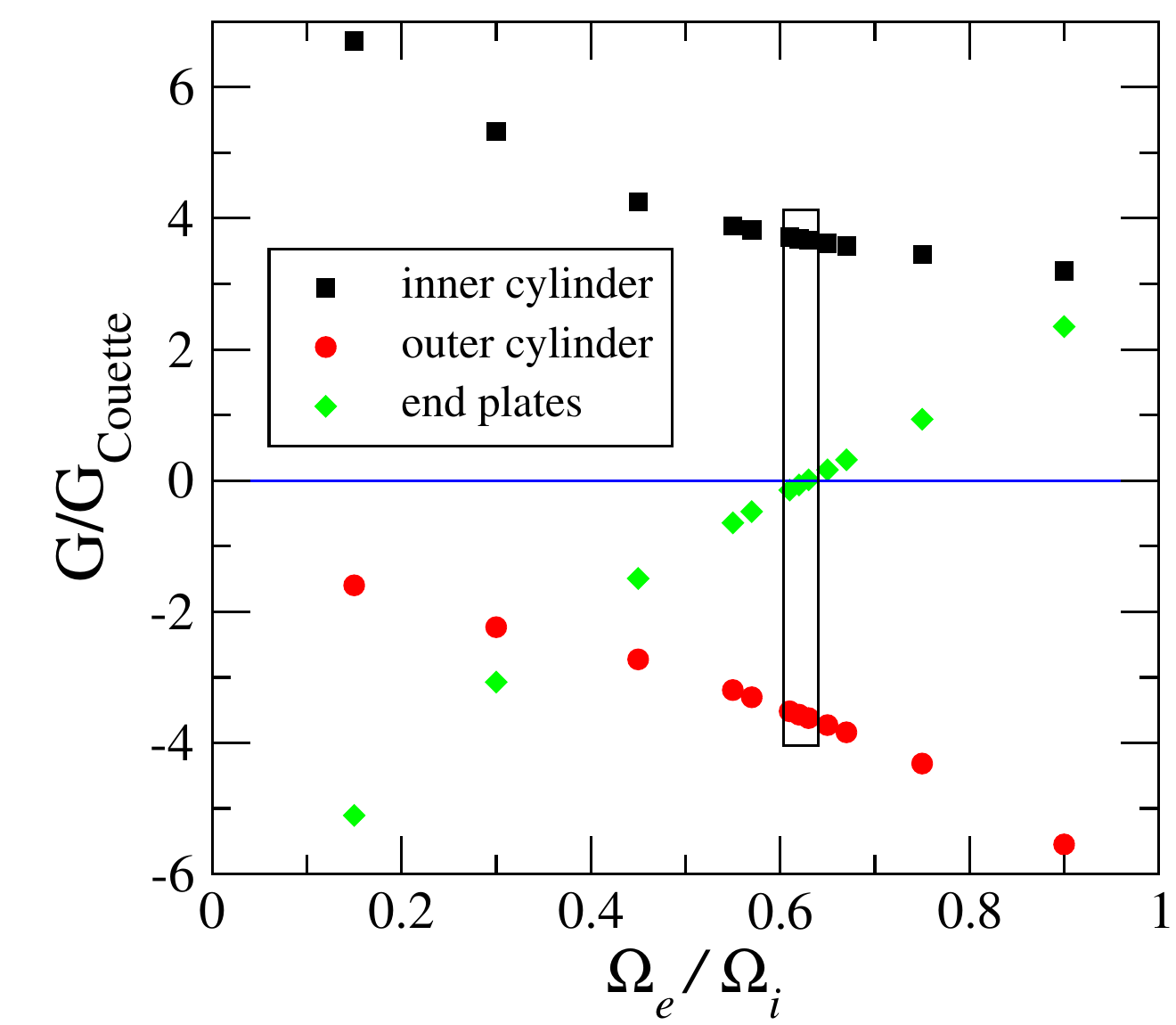} \\                                                   
    \end{tabular}                                                                                      
  \end{center}                                                                                         
  \caption{$(a)$ Rotation speed of the end plates ($\Omega_e$) normalised                              
    by the rotation speed of the inner cylinder ($\Omega_i$)                                           
    in the \emph{WR}, \emph{HTX} and \emph{MRI} configurations.                                        
    $(b)$ Torque ($G$) on the cylinders and end plates in the                                          
    the \emph{WR} configuration as a function of $\Omega_e$.                                           
    $G$ is normalised by the torque of the Couette profile~\eqref{Couetteflow}                         
    ($G_\text{Couette}$). The optimal rotation range is enclosed in a rectangular box.                                       
    $(c)$ Same as in $(b)$ for the \emph{HTX} configuration. In both $(b)$ and                         
    $(c)$ torques were computed at $R_s = 644$.}                                                       
  \label{axial_bounconds}                                                                              
\end{figure}

 Different configurations of this apparatus                                                   
differ from one another in the number of                                                               
rings into which the end plates are split.                                                           
Here, we study in detail two configurations,                                                           
the so-called \emph{HTX} and \emph{wide ring} (\emph{WR}),                                             
in which only one ring rotates differentially with respect                                             
to the cylinders. In the \emph{HTX} configuration                                                      
(dashed line in figure~\ref{axial_bounconds} $(a)$)                                                    
the end plates are split in three rings. The inner and outer rings                                     
are attached to the cylinders, whereas the central ring rotates                                        
at an angular velocity $\Omega_e$ intermediate to those of                                             
the cylinders. In the \emph{WR} configuration                                                          
(solid line in figure~\ref{axial_bounconds} $(a)$)                                                     
there is a single ring that spans the entire annulus but also rotates independently. The dot-dashed line
in figure~\ref{axial_bounconds} $(a)$ illustrates the boundary conditions of \cite{JiBuSchGo06}, who designed
their device to study the magnetorotational instability by using electrically conducting fluids and                 
an imposed axial magnetic field. We will refer to this configuration as \emph{MRI}.                    

\subsection{Numerical method}                                                            
                                                                                                                                                                  
 The Navier--Stokes equations have been solved in cylindrical coordinates using a second           
order time-splitting method \citep{HuRa98,MeBaAl10}. The spatial discretization              
is via a Galerkin-Fourier expansion in $\theta$ and Chebyshev                             
collocation in $r$ and $z$. Hereafter the radial $u$, azimuthal $v$ and axial $w$ velocities
are normalized with respect to the characteristic velocity $\tilde{S} d$ used in the definition
of the shear Reynolds number \eqref{control_params}. The code used is a parallelized version of
a spectral solver that has been widely tested~\citep{AGLM08,LoMa15,LoMaAv15}. Details of the 
parallelization strategy can be found in~\cite{ShiRaHoAv15}. 

 In the \emph{WR} and \emph{HTX} configurations there are discontinuities in the  angular velocity                 
at the junctions where elements rotating at different speeds meet.                        
For an accurate use of spectral techniques these discontinuities                          
must be regularised (see~\cite{LoSh98}). In the \emph{WR} configuration            
this is accomplished through the introduction of two exponential functions in the form    
\begin{equation}\label{Om_wide}
  \begin{split}
    \Omega(r) = \Omega_e+(\Omega_o - \Omega_e)e^{-(r-r_o)/\epsilon}\\       
    +(\Omega_i-\Omega_e)e^{(r-r_i)/\epsilon},
  \end{split}
\end{equation}                                                                            
whereas in the \emph{HTX} configuration two hyperbolic tangent functions are used         
\begin{equation}\label{Om_HTX}                                                            
  \begin{split}                                                                           
    \Omega(r) = 0.5(\Omega_i+\Omega_o + (\Omega_o - \Omega_e)\,\tanh((r-\hat{r_o})/\epsilon)\\
    +(\Omega_e-\Omega_i)\,\tanh((r-\hat{r_i})/\epsilon)),                               
  \end{split}                                                                             
\end{equation}                                                                            
\noindent where $\hat{r_o}=r_i+0.48$ and $\hat{r_i}=r_i+0.071$ are the radial locations
at which the central ring meets the outer and inner rings respectively.
In both cases $\epsilon = 0.01$ was used.       

\begin{figure}
  \begin{center}
    \includegraphics[width=0.5\linewidth]{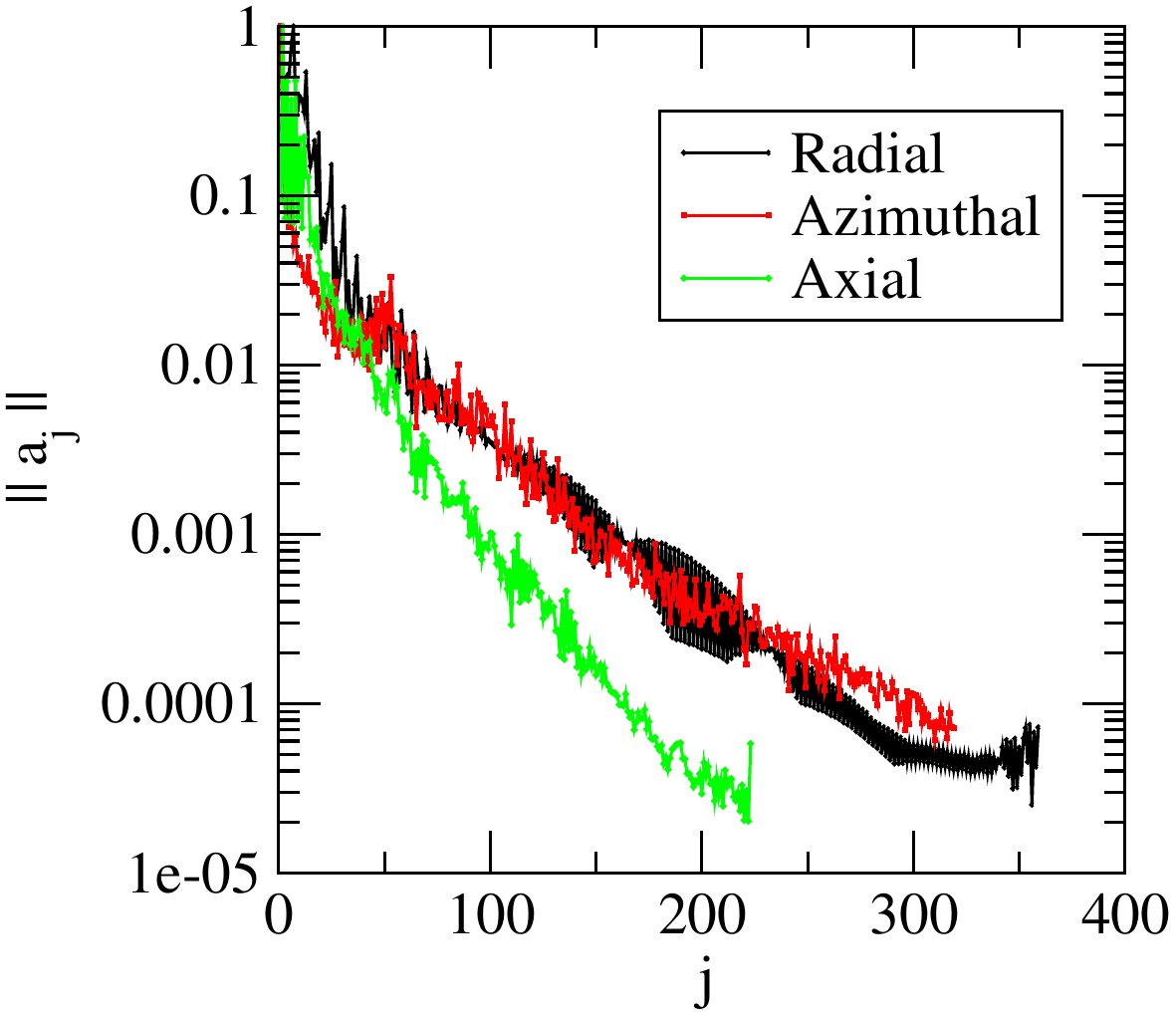}
  \end{center}
  \caption{Convergence of the spectral coefficients
     of the radial velocity $u$ in the three spatial
    directions using the infinity norm.
    It corresponds to a simulation of the \emph{HTX} configuration at $R_s= 25747$.
    This solution has been computed with $L=360$
    Chebyshev radial points, $M=320$ Fourier modes and $N=221$
    Chebyshev axial points.}
  \label{converge}
\end{figure}

The numerical resolution was carefully chosen in order to meet several requirements.
First, we checked that the total angular momentum flux through cylinders and end plates vanished in
the statistically stationary regime.
Second, we gradually increased the number of collocation points until
converged values of the torque at the inner and outer cylinders were obtained.
Finally, we checked the spectral
convergence of the code using the infinity norm of the spectral
coefficients of the computed solutions, defined as
$||a_l||_\infty=\max_{n,m}\,|a_{l,n,m}|$ for the radial direction, and
analogously for the axial and azimuthal directions. An example of
spatial convergence is illustrated
in figure~\ref{converge}, which shows $||a_j||_\infty$, with
$j=l,n,m$, of the radial velocity $u$ for a solution corresponding to
the \emph{HTX} configuration at $R_s=25747$. This solution was computed with
$L=360$ and $N=221$ Chebyshev axial points in $r$ and $z$,
and $M=320$ Fourier modes in $\theta$. In all simulations                                                        
the trailing coefficients of the spectral expansion                            
were at least four orders of magnitude                                                     
smaller than the leading coefficients.  
Table~\ref{res} shows the spatial resolution
corresponding to the largest $R_s$
simulated for each configuration. The reader is referred to~\cite{BrEck13}
for a comprehensive analysis on the suitability of
these convergence criteria for Taylor--Couette flows
at large Reynolds numbers.
\begin{table}
  \begin{center}
  \begin{tabular}{c c c c c}
    Configuration & $L$ &  $N$ & $M$ & $R_s$\\\hline
    \emph{WR}  & 128 & 256 & 544 & 47630\\
    \emph{HTX} & 392 & 642 & 224 & 32180\\
    \emph{MRI} & 192 & 192 & 288 & 12874\\\hline
  \end{tabular}
  \end{center}
  \caption{ Maximum spatial resolution and $R_s$ for
    each configuration. $L$,$N$ and $M$ indicate the number of spectral modes
    in the radial, azimuthal  and axial directions, respectively.}
  \label{res}
\end{table}

\subsection{Optimal boundary conditions}\label{sec:opt}                                               
                                                                                                      
Following~\cite{EdJi15}, we here determine the optimal rotation speed of the end plates     
from a balance of the angular momentum                                                                
fluxes (torque) through the boundaries of the apparatus.                                              
In particular, optimal rotation is identified when                                                    
the torque at the end plates ($G_e$) becomes zero,                                                    
so that the torque on the cylinders                                                                   
has the same magnitude but opposite sign,                                                             
$G_i = -G_o$, as in the infinite-cylinder idealization.                                               
Figures~\ref{axial_bounconds} $(b)$ and $(c)$  show the torque across the cylinders and end plates    
as a function of $\Omega_e$ for solutions computed                                                    
at $R_s = 644$ in both configurations. In agreement with~\cite{EdJi15},                               
in both cases there exist a narrow                                                                    
range of $\Omega_e$ for which $G_e = 0$ is                                                            
approximately fulfilled                                                                               
(rectangular box in figures~\ref{axial_bounconds} $(b)$ and $(c)$).                                   
For our simulations we chose                                                                          
$\Omega_{e} = 0.62 \Omega_i$ (\emph{HTX}) and $\Omega_{e} = 0.21 \Omega_i$ (\emph{WR}).               
Note that the torque in the \emph{WR}                                                                 
configuration is substantially larger than                                                            
in the \emph{HTX} setup. The reason for this will                                                     
be discussed in \S\ref{sec:low_Re}, along with the description                                  
of the secondary flows in both setups.

\section{Basic flow and transition to turbulence}\label{sec:low_Re}            

\subsection{Wide ring}\label{sec:low_wide}                                             

\begin{figure}
  \begin{center}\setlength{\piclen}{0.2\linewidth}
    \begin{tabular}{cccc}
      \multicolumn{2}{c}{$(a)$}  & \multicolumn{2}{c}{$(b)$} \\
      \includegraphics[width=\piclen]{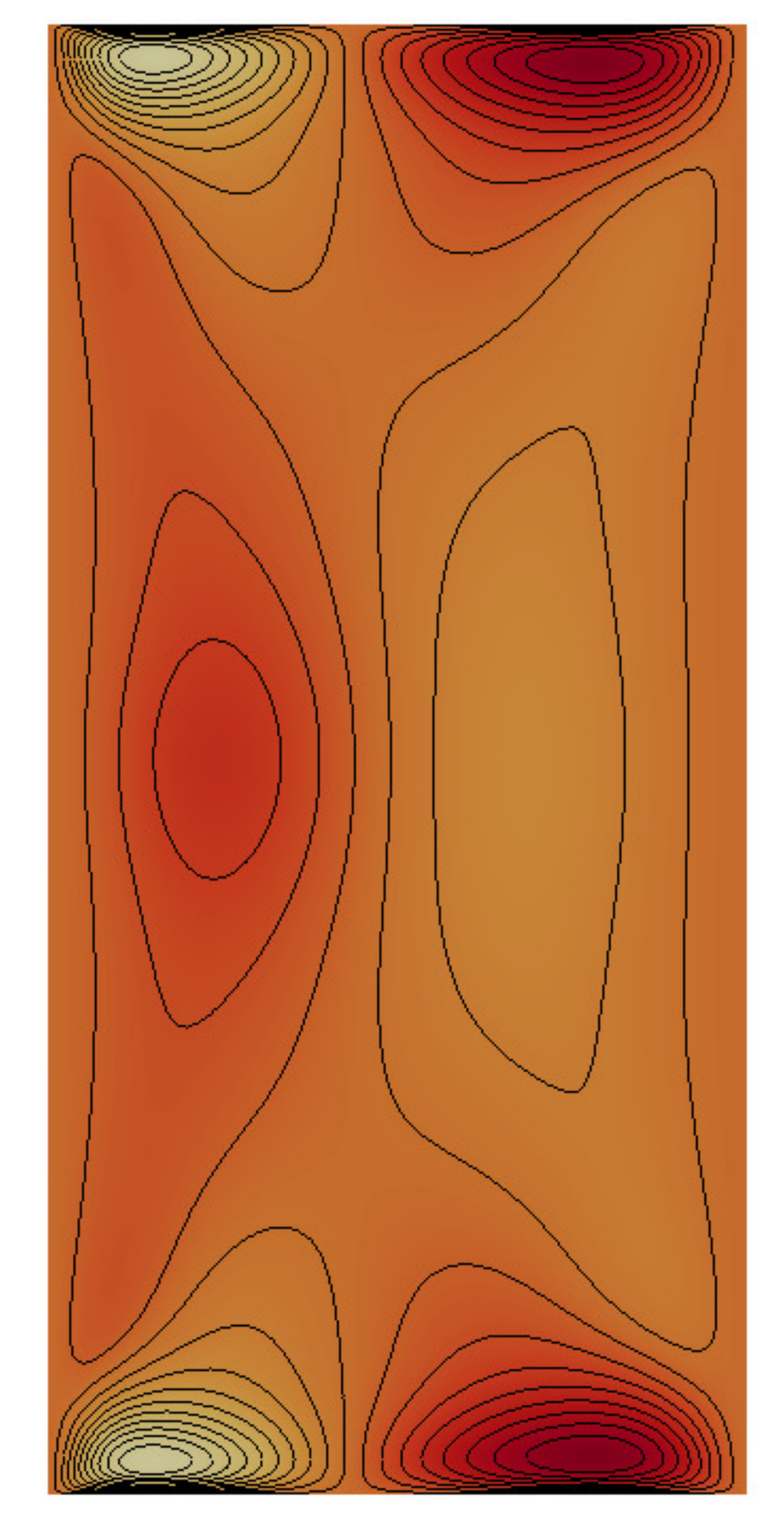} &
      \includegraphics[width=\piclen]{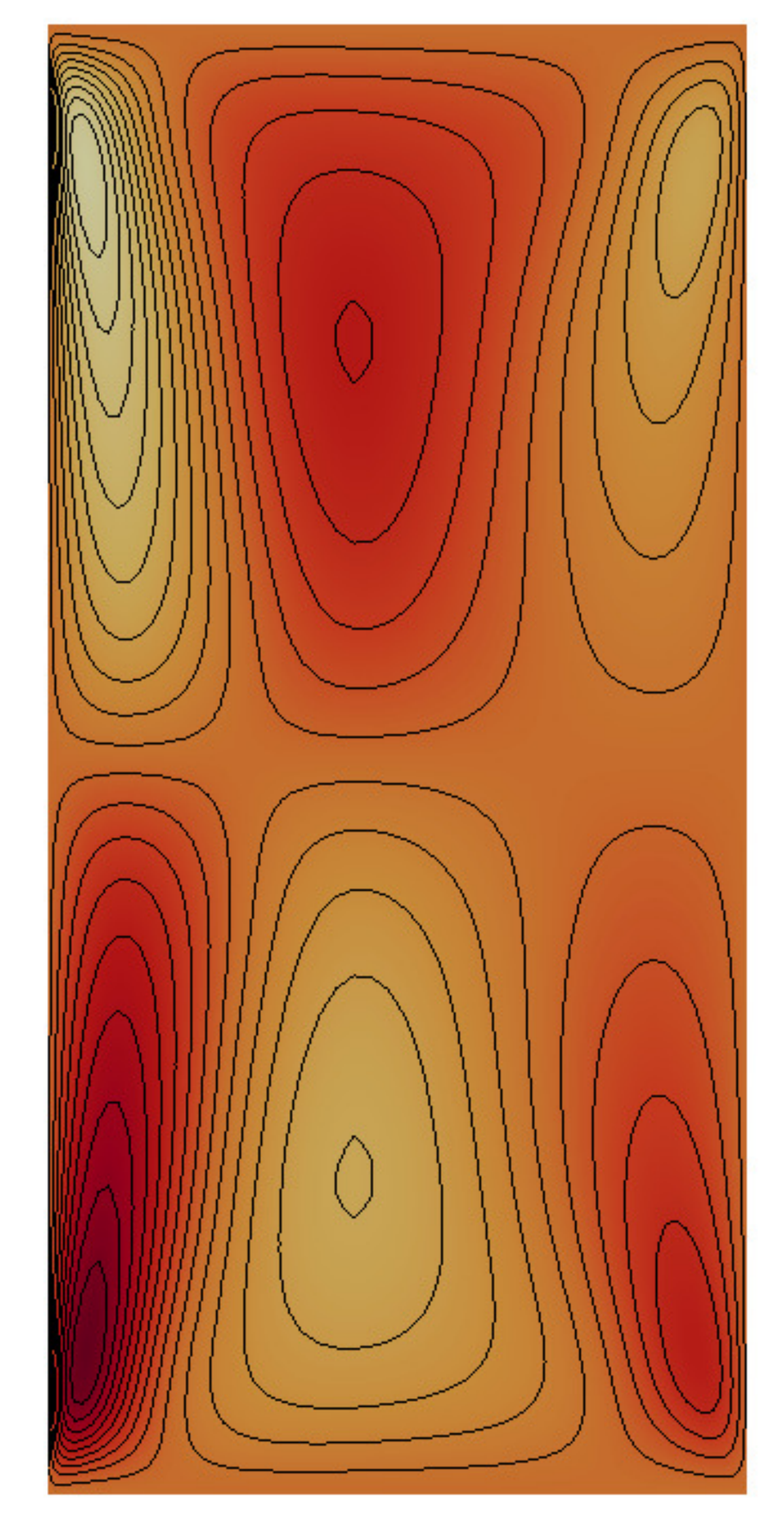}&
      \includegraphics[width=\piclen]{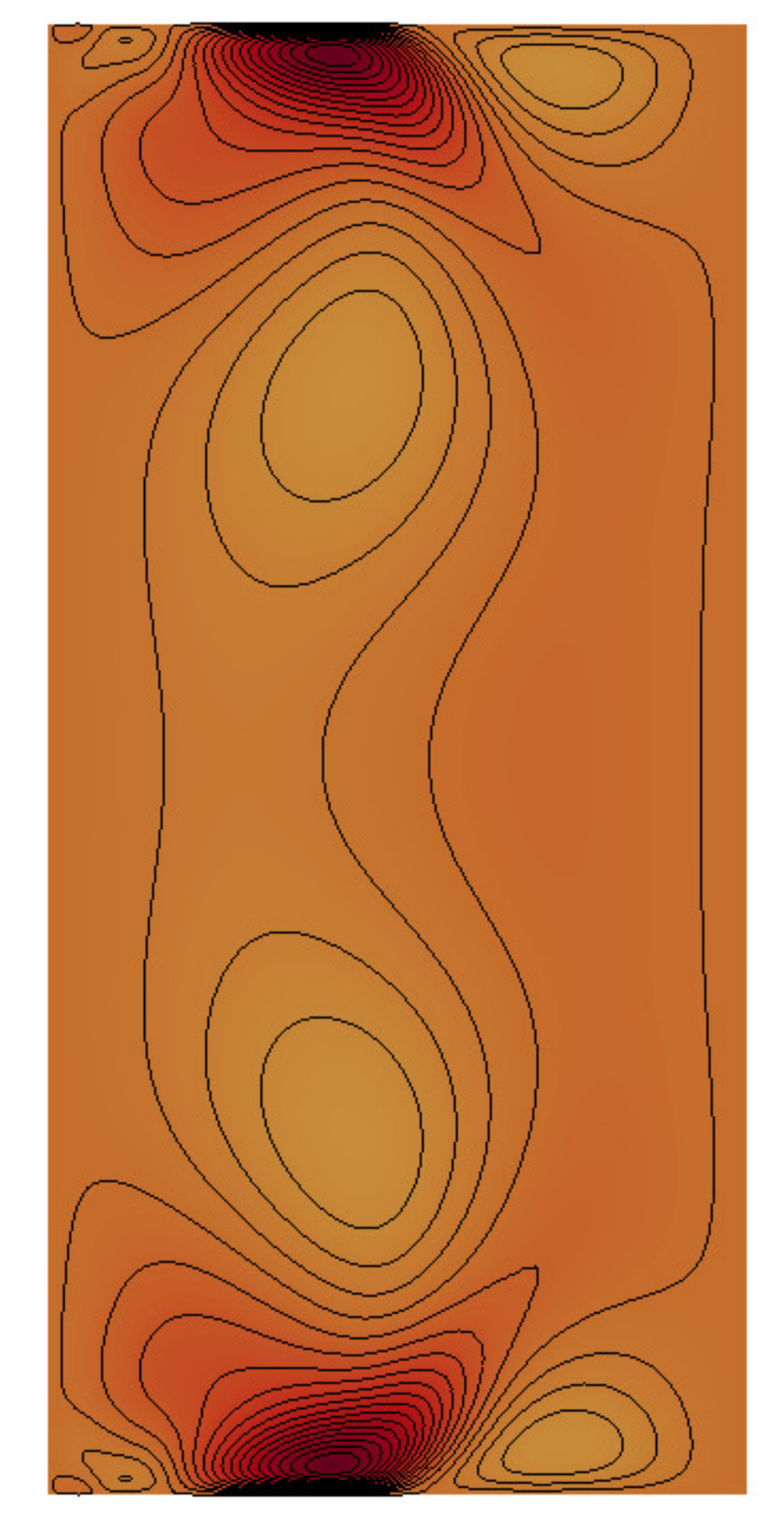} &
      \includegraphics[width=\piclen]{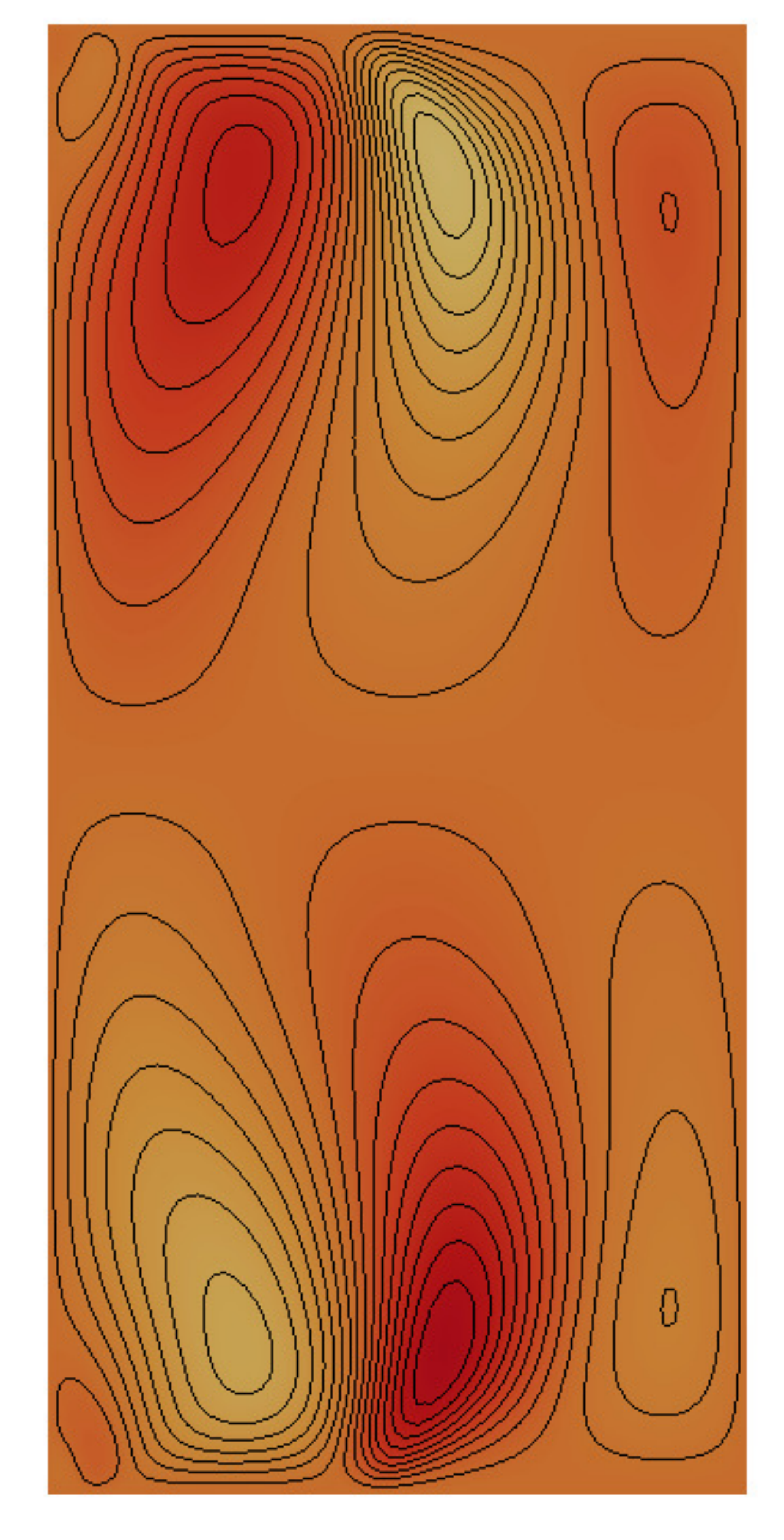} \\
      $u$ & $w$ & $u$ & $w$\\
    \end{tabular}
  \end{center}
  \caption{ Meridional sections $(r,z)$ showing color maps of
    the radial (left panel) and axial (right panel) velocities in the \emph{WR} $(a)$ and
    \emph{HTX} $(b)$ configurations. The solutions depicted correspond in both cases to laminar flow
    computed at $Re_s = 644$. In all figures shown in this paper, dark (light) regions
    indicate zones of positive (negative) velocities. There are $20$ contours equally distributed  
    in $u \in [-0.08,0.06]$, $w \in [-0.11,0.11]$ and $u \in [-0.02,0.08]$, $w \in [-0.07,0.07]$
    for the \emph{WR} and \emph{HTX} configurations respectively.}
  \label{meridional}
\end{figure}

 Figure~\ref{meridional} $(a)$ shows that in the \emph{WR} configuration 
the secondary EC cells extend                                                               
over the entire annulus.                                  
Near the end plates, the flow is deflected radially towards                               
the cylinders leading to two                                                     
Ekman vortices with opposite sense of                                                     
circulation. The size and intensity of these                                              
vortices change with  $\Omega_e$                                                          
and it is only under optimal boundary conditions that                                     
these have nearly equal size                                                              
and strength ($\max |u| = 0.08$).
When the fluid reaches the cylinders                                        
it is transported from the end plates to the mid-plane over                               
Stewartson boundary layers. As a result,
two strong radial jets emerge from the cylinders and                                      
displace the flow towards mid-gap.  The circulation                                       
cycle is then closed by two vertical cells                                                
that transport the fluid back to the end plates.                                          
                                                                                          
 These large-scale secondary flows
lead to linear instabilities that
manifest themselves at the equatorial region~\citep{AGLM08,Av12},
and cause a transition to turbulence at low values of
$R_s$. Figure~\ref{eigenandturbulent} $(a)$ shows, through isosurfaces of the
radial velocity $u$, the location of the most unstable mode
at the onset of instability. Note that
the axisymmetric part of $u$ has been subtracted
to facilitate visualization. The flow pattern emerging from
this primary transition, which takes place at $R_s \approx 865$,
is a rotating wave with azimuthal wave number $m=2$. As $R_s$
is further increased, the flow undergoes secondary instabilities
leading to either rotating waves with different $m$ or quasi-periodic
states, and becomes eventually turbulent at $R_s \approx 3218$.
Figure~\ref{eigenandturbulent} $(c)$ shows isosurfaces of $u$
for a turbulent state computed at $R_s=3862$.
Interestingly, the turbulence does not extend towards the end plates
but remains concentrated around the mid-plane.

\subsection{HTX}\label{sec:low_HTX}                                                    
                                                                                          
 Figure~\ref{meridional} $(b)$ shows that in the \emph{HTX} configuration 
the secondary EC is mainly confined to the vicinity of the end plates.                                                  
Because of the split end plates, the radial flow along them                          
is arranged in four alternating outward-inward vortices                                   
which direct the flow towards the junctions between the rings.                                                                                        
The pair of vortices located at the outermost part of                                     
the end plates are significantly larger and more intense ($\max |u| = 0.08$)                                  
than those arising near the inner cylinder ($\max |u|  = 0.02$).                                      
Significant axial transport of fluid towards                                 
the mid-plane  occurs  only                                                    
in a narrow region around mid-gap.  However,                                              
the flow does not reach the mid-plane,                                                    
as it is recirculated towards the inner cylinder                                          
by a strong radial inflow that arise  at an approximately intermediate                    
distance between the end plates and                                                       
the equatorial region. Finally,                                                           
the flow is pushed back towards                                                           
the end plates by axial velocities                                                        
arising in the regions near the cylinders.
A comparison with the \emph{WR} configuration reveals that
the substantial difference in torque                                  
shown in section~\ref{sec:opt} is caused by the influence of                              
the Stewartson boundary layers that form at the cylinders in                                         
the \emph{WR} configuration. These produce strong                                  
azimuthal velocity gradients near the cylinders,                                          
which result in a significant increase of the torque                   
as compared with that in the \emph{HTX} configuration.
                                    
\begin{figure}\setlength{\piclen}{0.22\linewidth}                                                                            
  \begin{center}                                                                        
    \begin{tabular}{cccc}                                                                   
      $(a)$  & $(b)$ & $(c)$ & $(d)$\\                                                                   
      \includegraphics[width=\piclen]{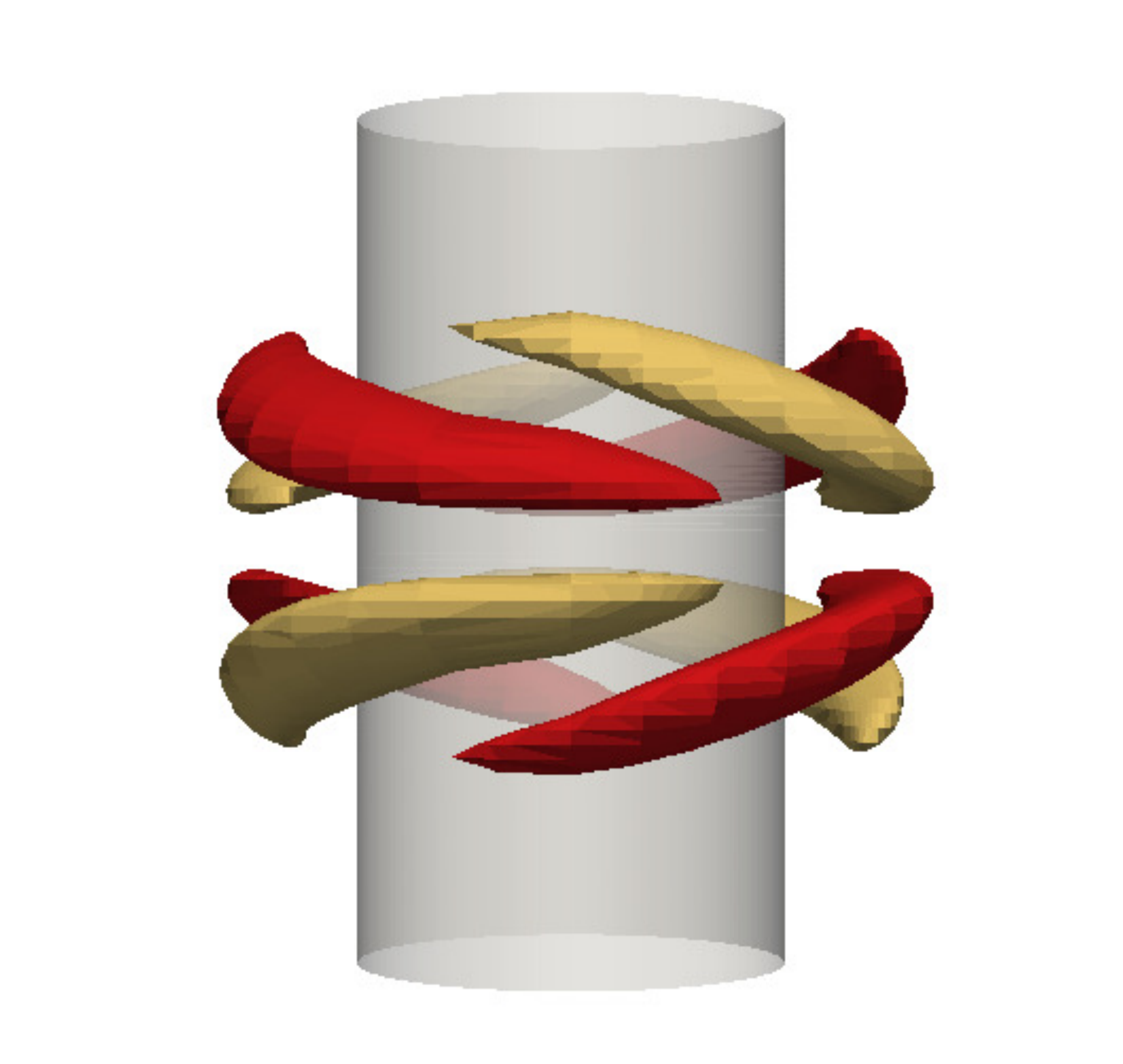} &                           
      \includegraphics[width=\piclen]{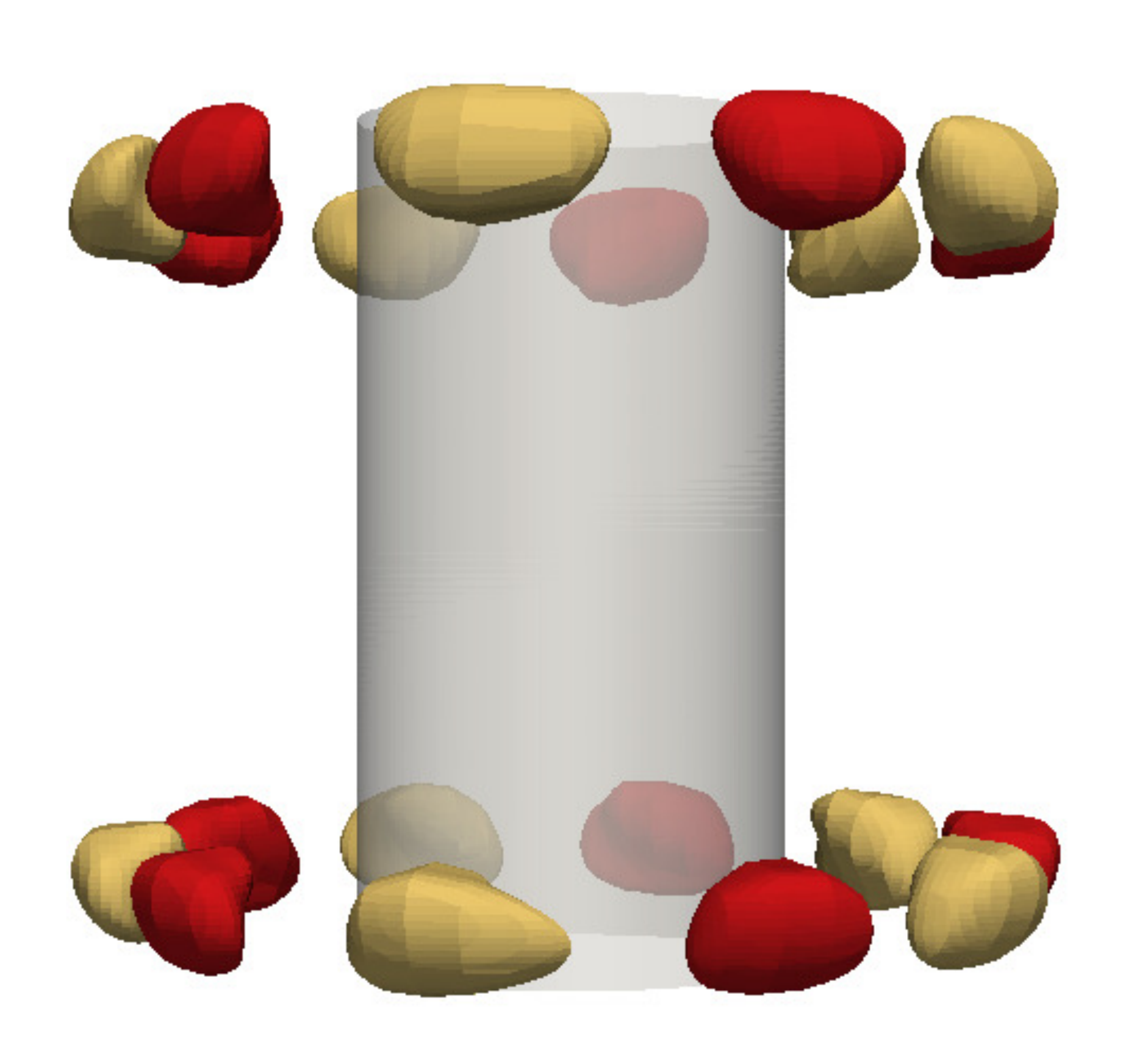}&                            
      \includegraphics[width=\piclen]{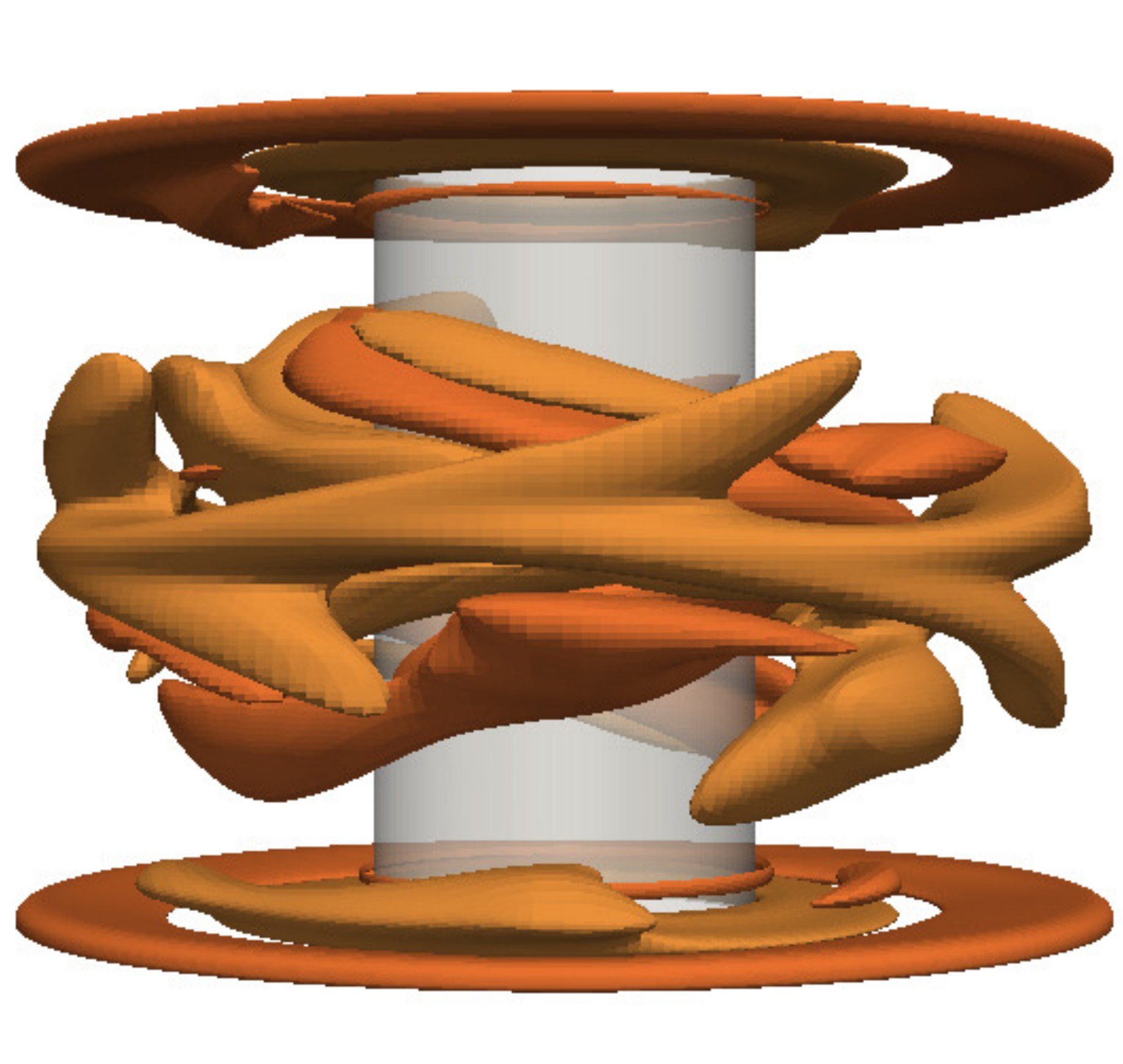} &
      \includegraphics[width=\piclen]{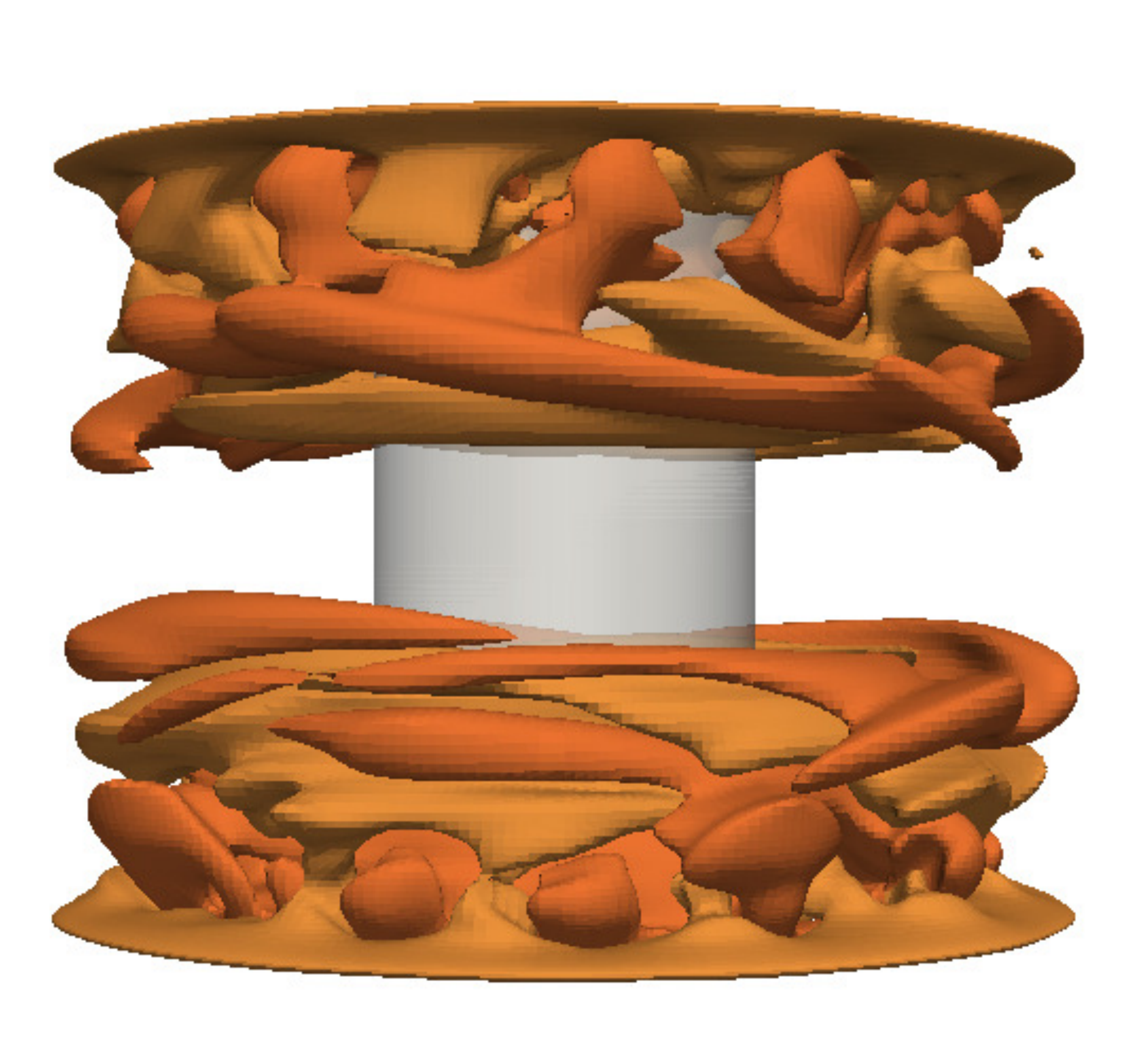}\\
    \end{tabular}                                                                         
  \end{center}                                                                            
  \caption{$(a)$ and $(b)$ show isosurfaces of the radial velocity of
    the leading eigenmodes near the onset of instability.
    The axisymmetric component has been subtracted to facilitate visualization of the spatial structure
    of the unstable mode. $(a)$ Equatorial rotating wave with $m=2$ computed at $R_s=772$ in the \emph{WR} configuration, $(b)$
    Rotating wave with $m=5$  localised near the end plates for $R_s=901$ 
    in the \emph{HTX} configuration. $(c)$ and $(d)$ show isosurfaces of the radial velocity for
    turbulent states computed at $R_s=3862$ in the \emph{WR} $(c)$ and
    \emph{HTX} $(d)$ configurations.  There are $6$ isosurfaces equally distributed 
      across $u \in [-0.13,0.12]$ and $u \in [-0.09,0.14]$ in the \emph{WR} and
      \emph{HTX} configurations respectively.}
  \label{eigenandturbulent}                                                                           
\end{figure}

 The meridional circulation in the HTX configuration becomes unstable
at $R_s \approx 727$. The instability results in a rotating wave                                    
with $m=5$ localised at the end plates (see figure~\ref{eigenandturbulent} $(b)$),
and the flow becomes quickly chaotic  as $R_s$ is increased ($R_s \approx 1287$).                    
Nevertheless, the turbulence remains primarily localised                                 
near the upper and lower third of the experiment (see figure~\ref{eigenandturbulent} $(d)$),        
so that the zonal flow at the equatorial region is barely                                 
affected by the secondary flows and nearly matches a quasi-Keplerian velocity profile,
 see \S\ref{sec:high_HTX}.

\section{Dynamics at high Reynolds numbers}

 As the rotation of the cylinders is increased                                            
the spatial arrangement of the secondary flows                                            
undergo significant changes in both configurations,
which alter the structure of the resulting turbulence. 
While this transition occurs smoothly with increasing $R_s$,
we here distinguish between low and high Reynolds numbers
using $R_s \approx 10^4$ as an approximate threshold,
beyond which the changes described in this section begin
to become apparent.

\subsection{Wide ring}                                                                 
                                                                                          
\begin{figure}\setlength{\piclen}{0.22\linewidth}                                                                                   
  \begin{center}                                                                                                                       
    \begin{tabular}{cccc}                                                                                                                
      \multicolumn{2}{c}{$(a)$} & \multicolumn{2}{c}{$(b)$} \\                                                                     
      \includegraphics[width=\piclen]{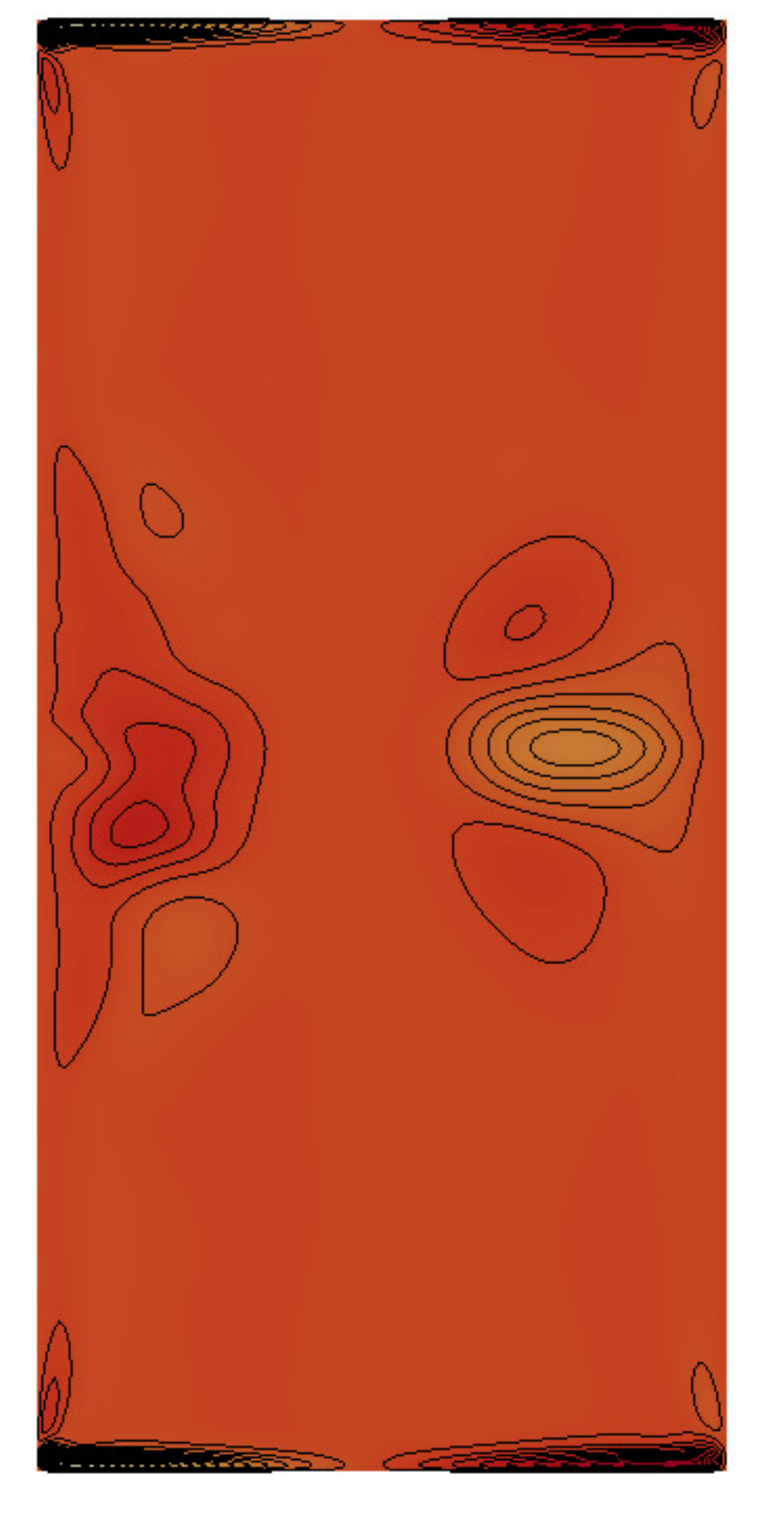} &                                                                       
      \includegraphics[width=\piclen]{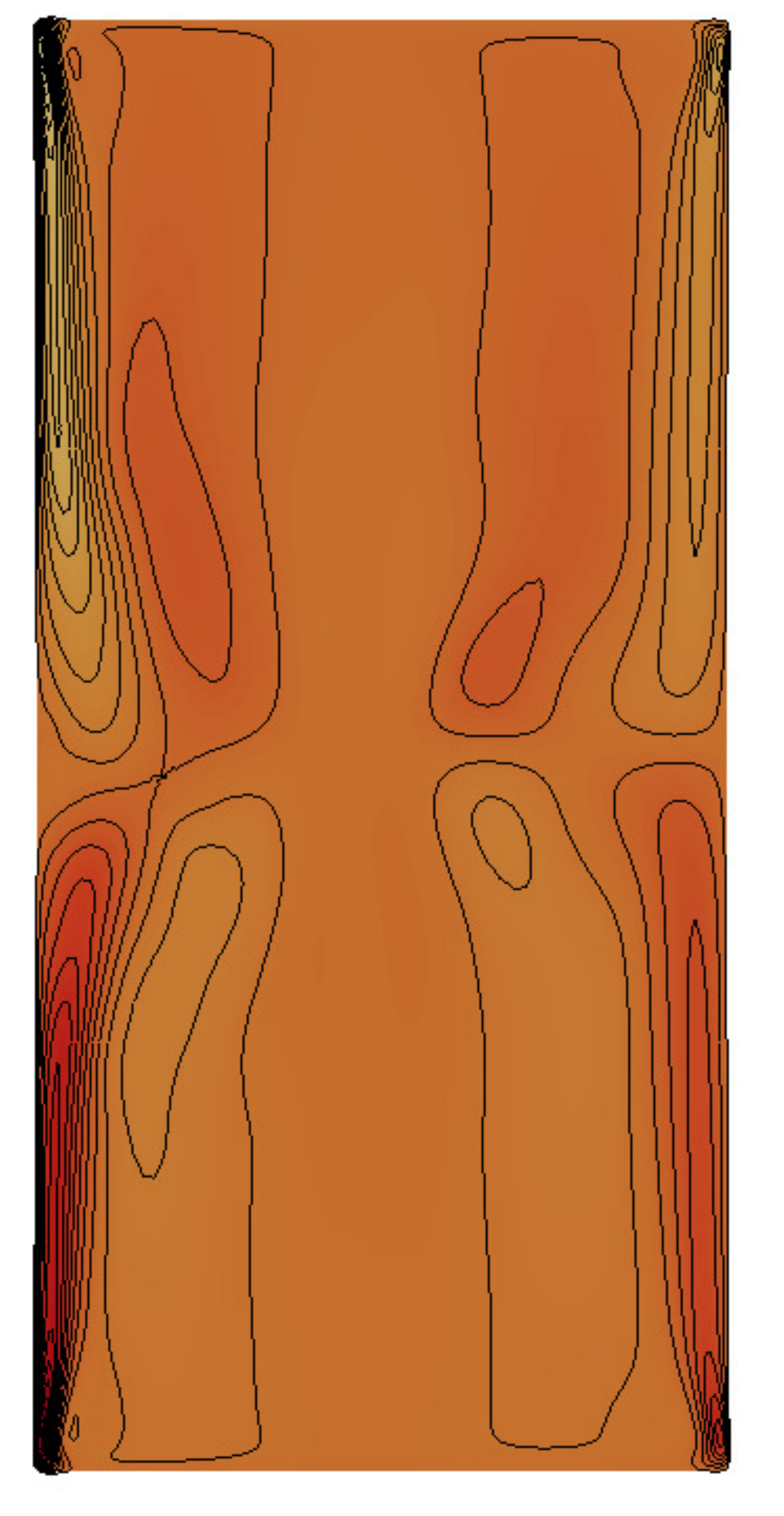}&
      \includegraphics[width=\piclen]{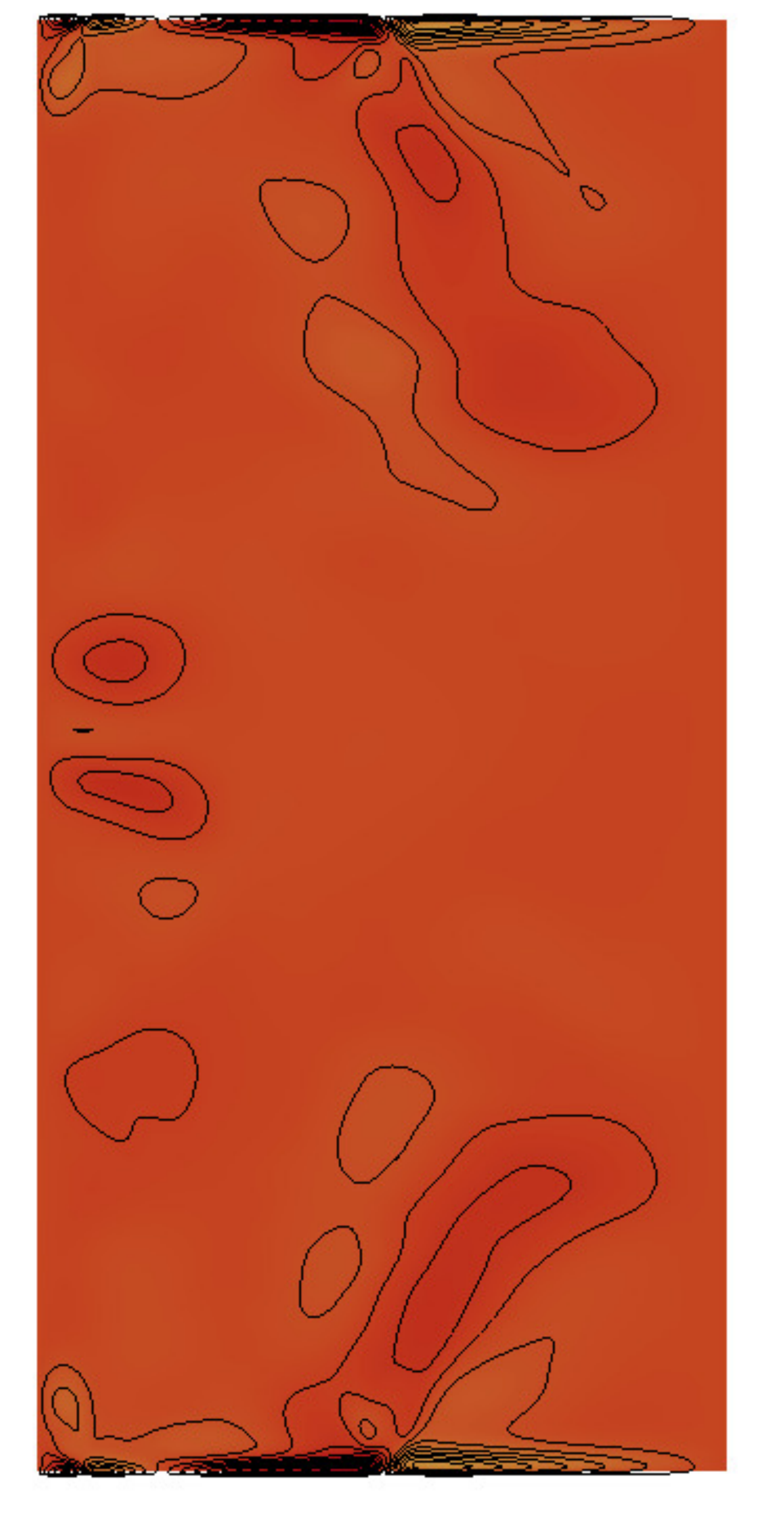} &
      \includegraphics[width=\piclen]{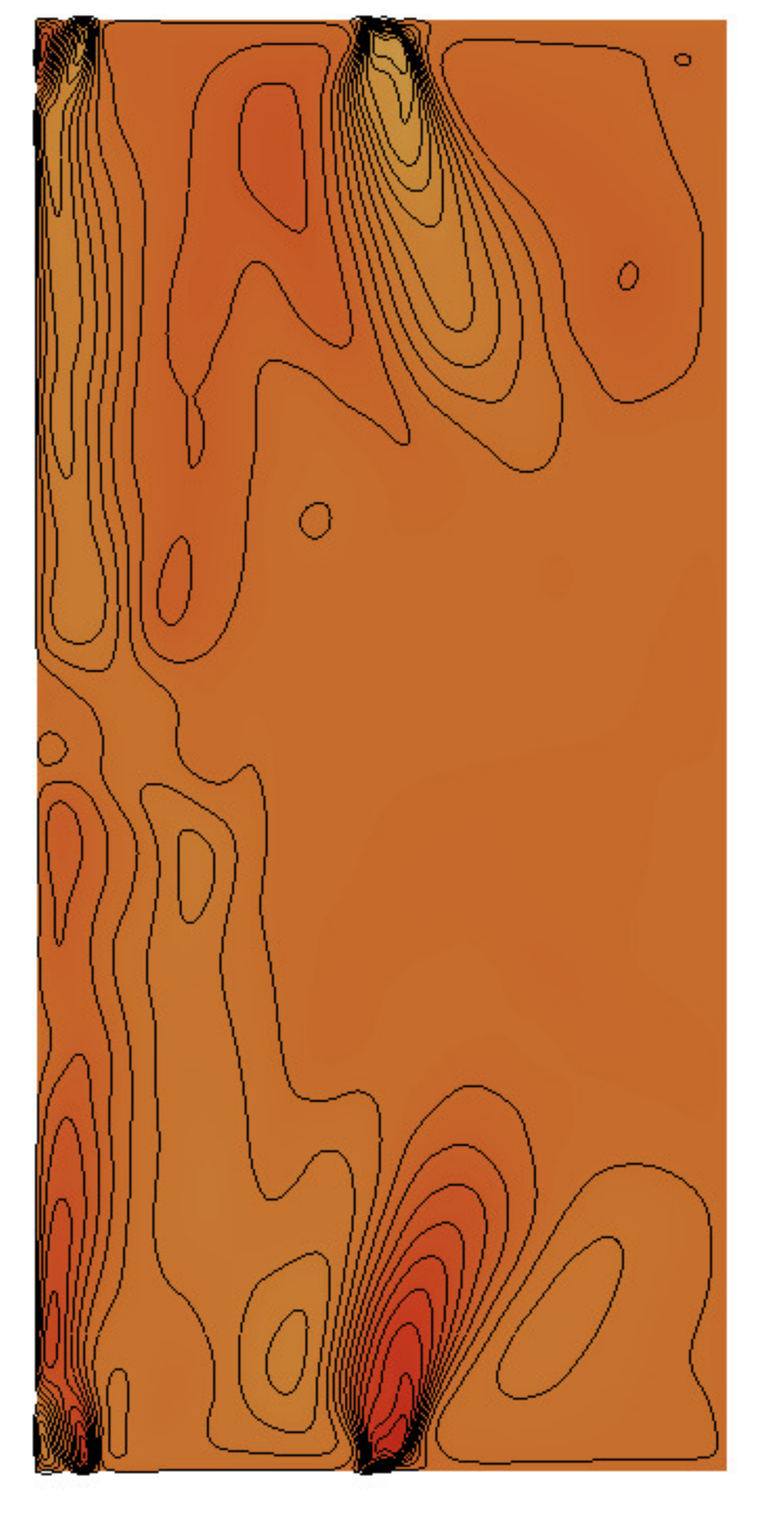} \\
      $u$ & $w$ & $u$ & $w$\\                                                                                                                      
    \end{tabular}                                                                                                                      
  \end{center}                                                                                                                         
  \caption{ Meridional sections $(r,z)$ showing color maps of the mean radial (left panel) and axial (right panel) velocities in        
    the  \emph{WR} $(a)$ and \emph{HTX} $(b)$ configurations. The solutions depicted were computed at $R_s = 19310$ in                 
    $(a)$ and $R_s=22529$ in $(b)$.  There are $20$ contours equally distributed                                 
      in $u \in [-0.16,0.08]$, $w \in [-0.17,0.17]$ and $u \in [-0.07,0.10]$, $w \in [-0.06,0.06]$
      for the \emph{WR} and \emph{HTX} configurations respectively.}                                                         
  \label{meridional_high}                                                                                                              
\end{figure}

 Figure~\ref{meridional_high} $(a)$  shows the structure
of the time-averaged secondary flow for the \emph{WR} configuration at $R_s=19302$. Here                                                  
the radial jets that emanate from the cylinders                                           
at the equatorial region do not extend across the entire gap,                             
but remain localised in regions closer to the cylinders.                                  
The gradual displacement of these                                                         
jets towards the cylinders as $R_s$ increases                                             
is accompanied by the emergence of                                                        
two pairs of radial flow cells on top and                                                 
bottom of them. These radial cells                                                        
recirculate the flow towards the cylinders,                                               
so that the vertical transport of                                                         
fluid from the equator towards the end plates                                             
that closes the Ekman circulation cycle is also confined                                  
to the vicinity of the cylinders.                                                         
As a result, the radial and axial velocities are nearly zero over the                     
central region of the gap and                                                        
the flow becomes essentially azimuthal in the bulk, resulting in vanishing Reynolds stresses.                                              

\begin{figure*}\setlength{\piclen}{0.33\textwidth}                                                                        
  \begin{center}                                                                                                          
    \begin{tabular}{ccc}                                                                                                  
        &  $(a)$ \emph{WR} &  \\                                                                                          
      \includegraphics[width=\piclen]{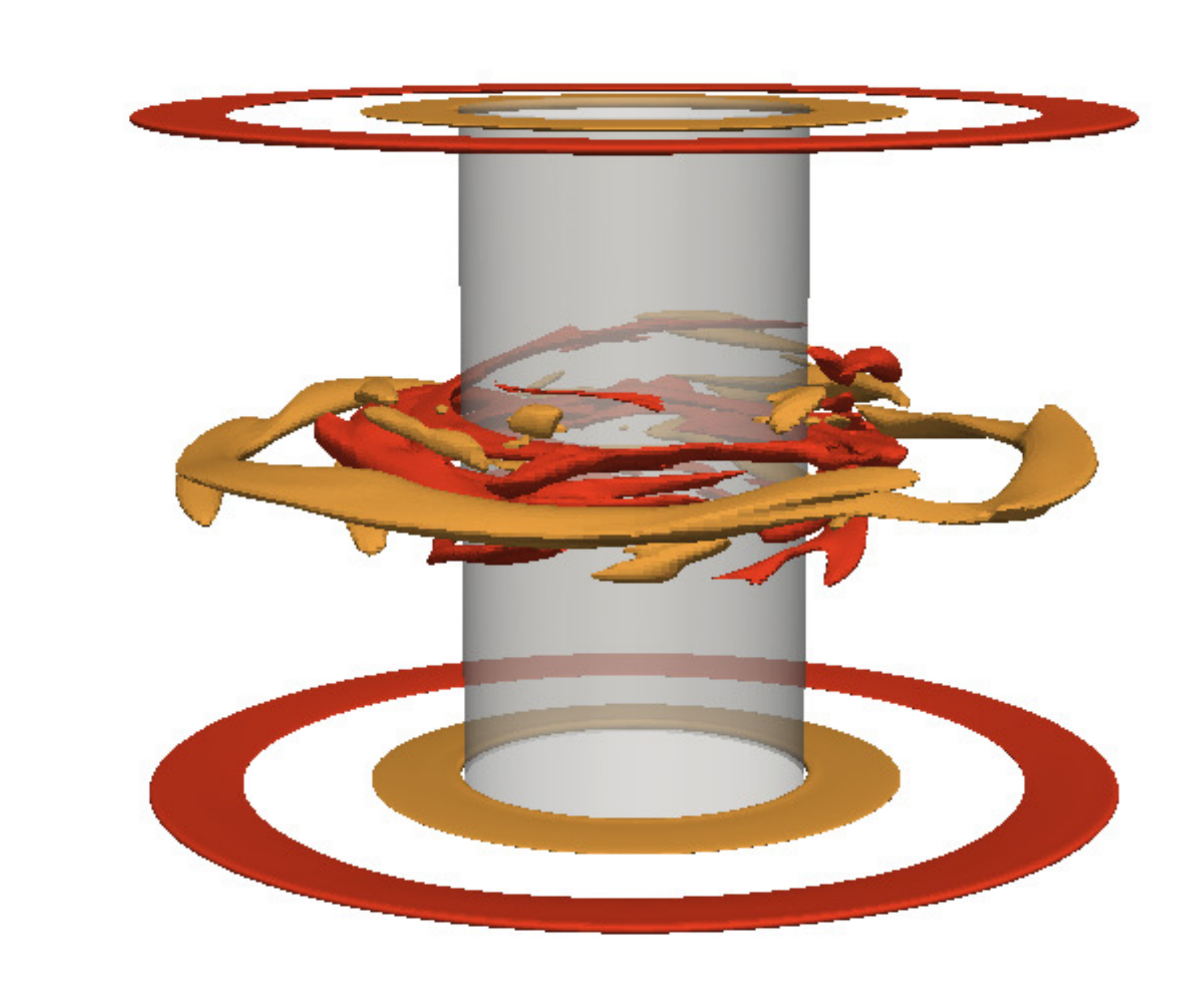} &                                                           
      \includegraphics[width=\piclen]{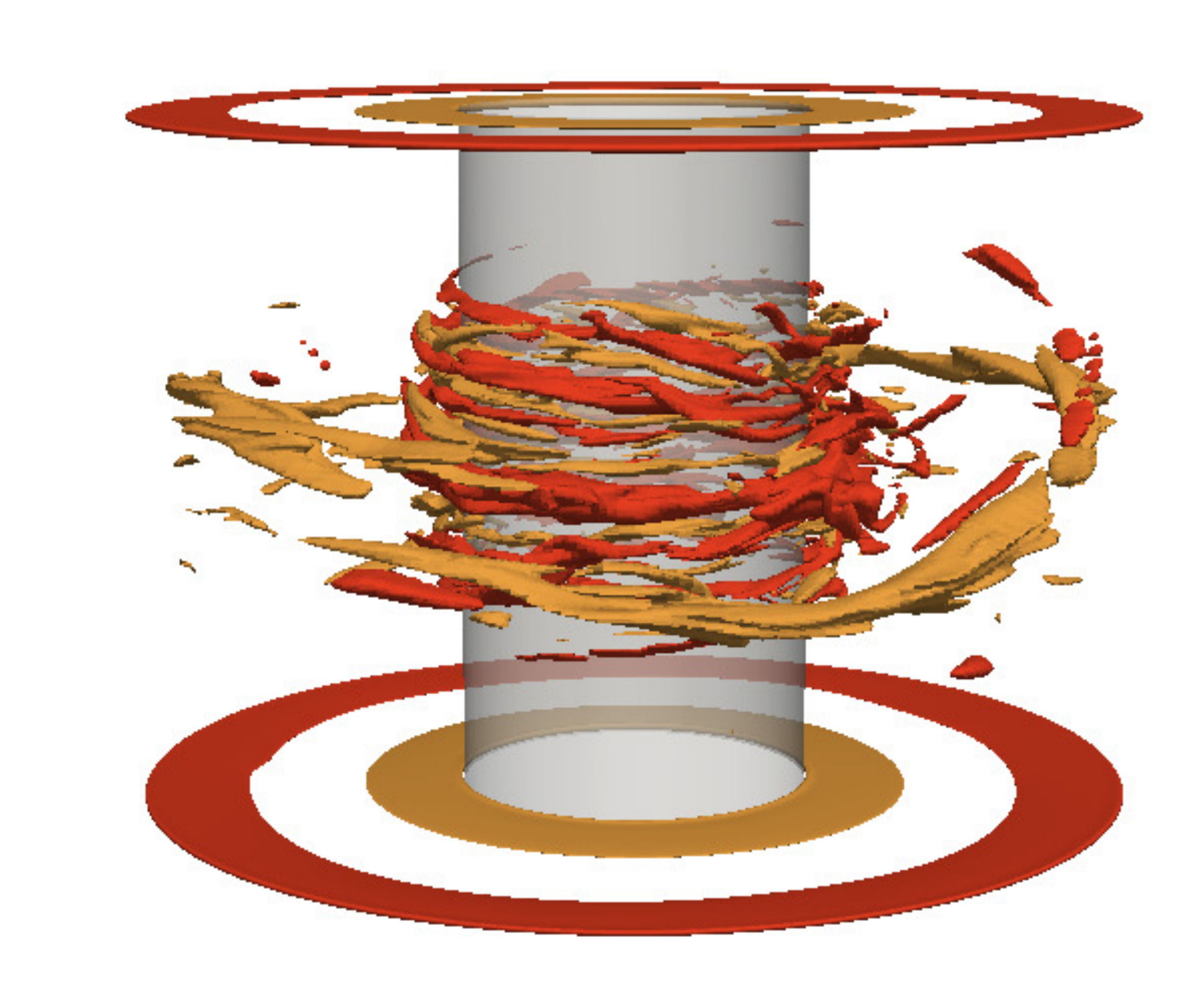}&                                                            
      \includegraphics[width=\piclen]{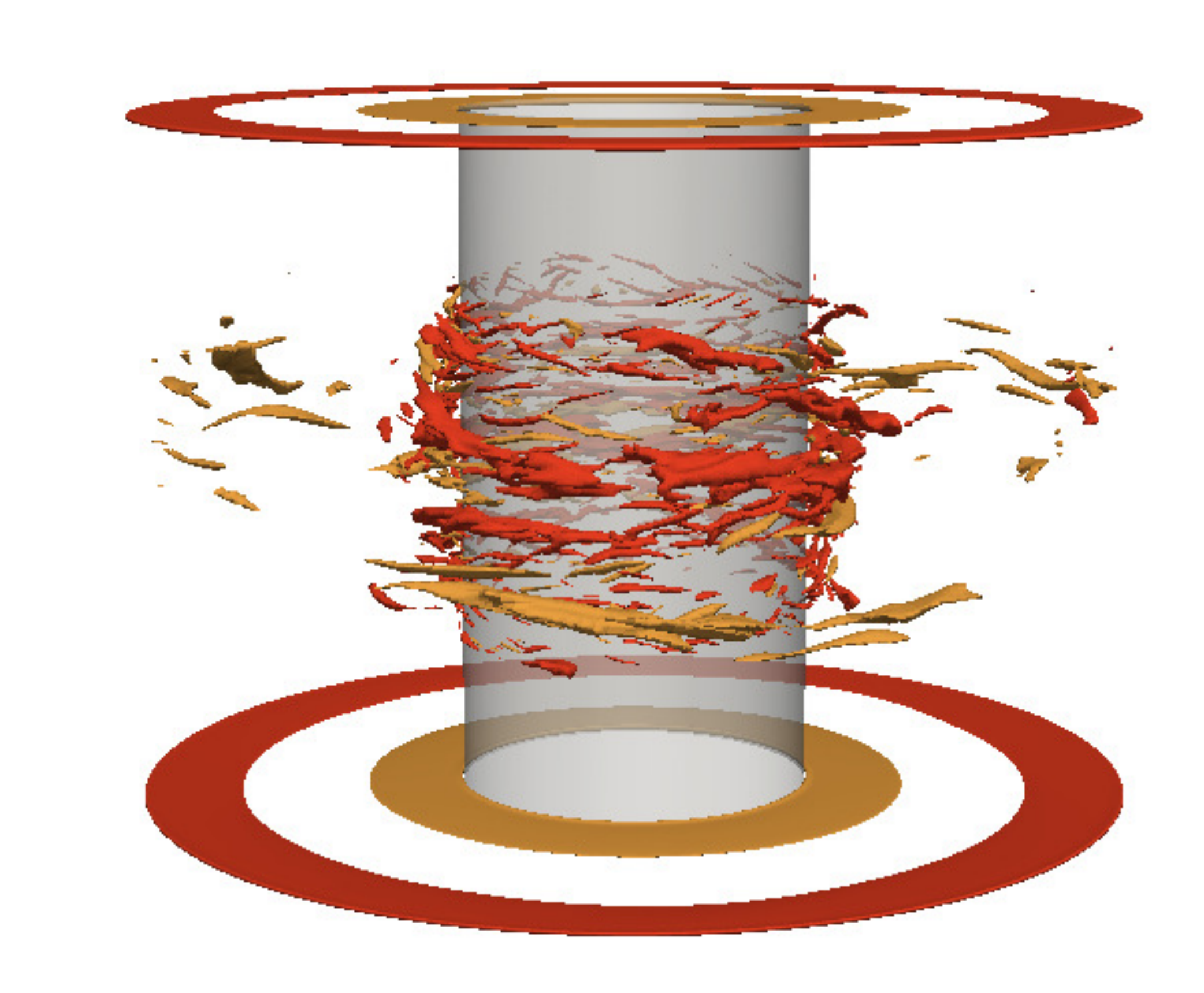} \\                                                          
      $R_s=11586$  & $R_s=27035$ & $R_s=47630$ \\                                                                         
        & $(b)$ \emph{HTX} &  \\                                                                                          
      \includegraphics[width=\piclen]{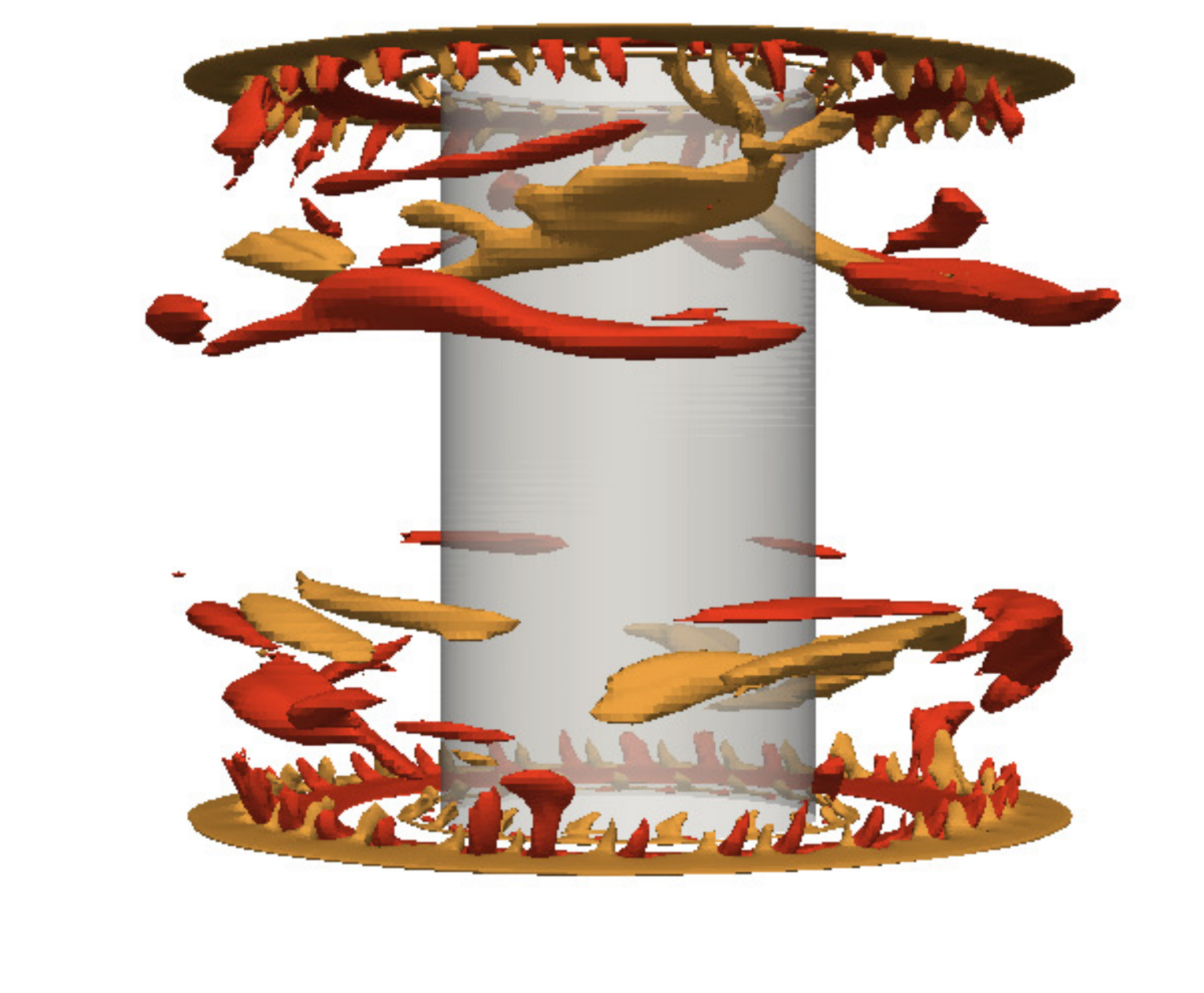} &                                                             
      \includegraphics[width=\piclen]{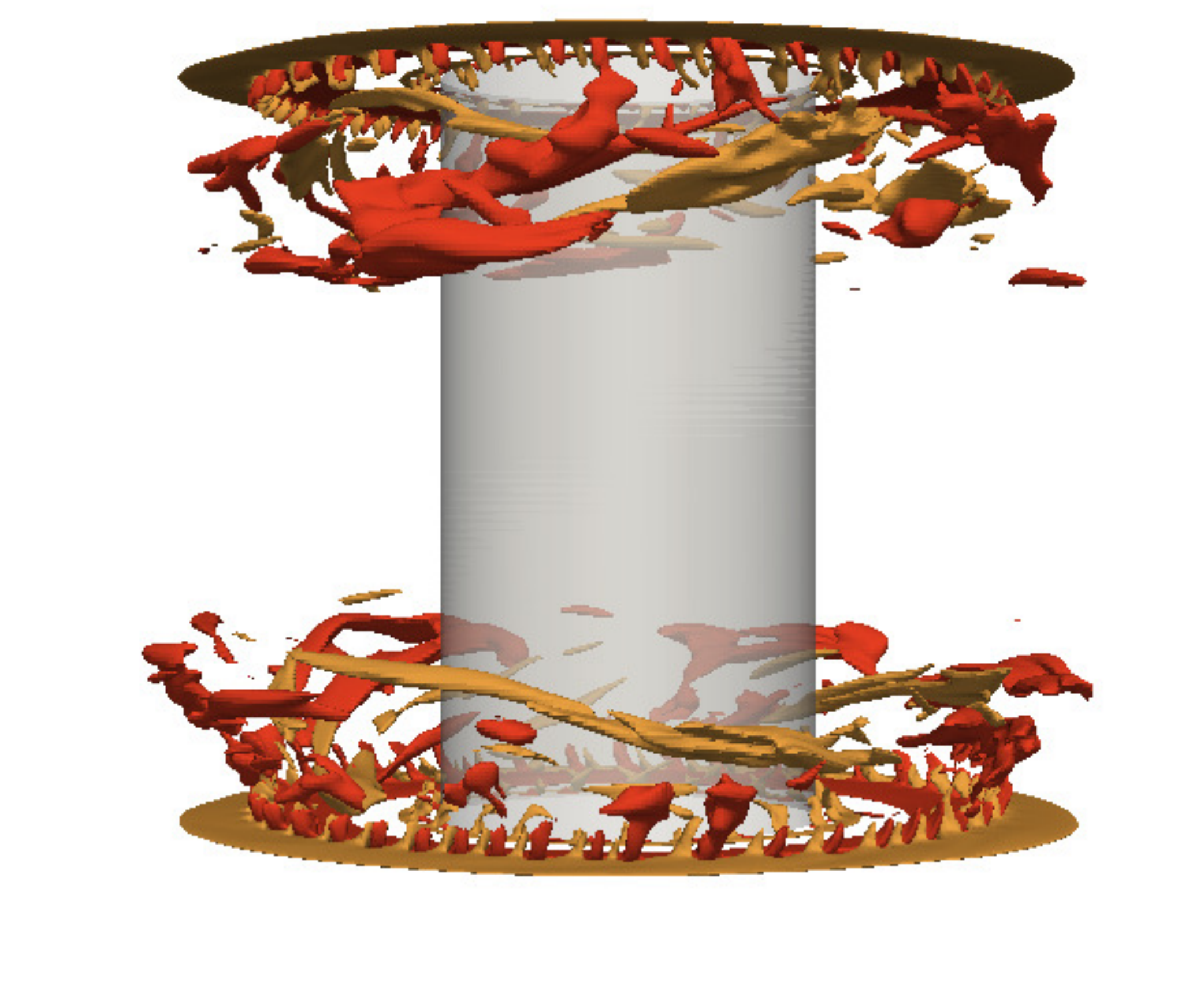}&                                                             
      \includegraphics[width=\piclen]{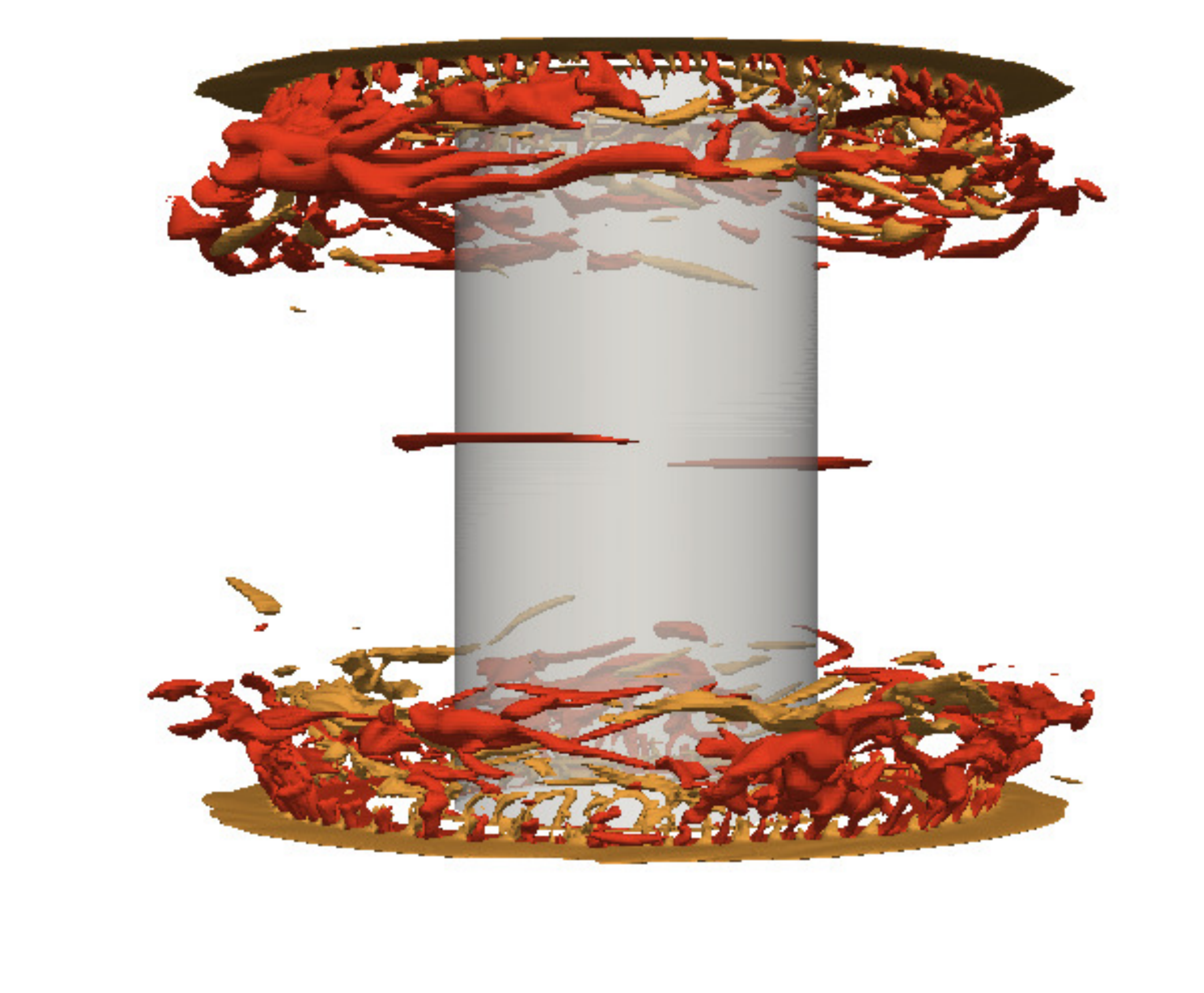} \\                                                           
      $R_s=9655$  & $R_s=19310$ & $R_s=32180$ \\                                                                          
                                                                                                                          
    \end{tabular}                                                                                                         
  \end{center}                                                                                                            
  \caption{ Evolution of the isosurfaces of the radial velocity $u$ with $R_s$ in both configurations.                    
    There are $2$ isosurfaces corresponding to $u= \pm 0.04$ in both cases. Note that plots corresponding to the \emph{WR} configuration            
    are slightly tilted to better illustrate the gradual displacement of the turbulence towards the cylinders.}           
  \label{turb_evol}                                                                                                       
\end{figure*}

 The turbulent dynamics of this system is confined to
the region in which the radial equatorial jets penetrate into the bulk flow. Hence  as $R_s$ increases and the secondary flows occupy regions closer to
the cylinders, significant turbulent fluctuations are only
found in the vicinity of the cylinders.
Figure~\ref{turb_evol} $(a)$ clearly illustrates
the progressive localisation of the turbulence
near the cylinders as $R_s$ increases. Interestingly,
turbulent structures occur mainly near
the inner cylinder, which could be related
to the large curvature of the apparatus ($\eta=r_i/r_o=0.3478$).

 Figure~\ref{profiles_wide} $(a)$ shows
the mean azimuthal velocity $v$ for
$R_s=47630$. With the exception of the zones
near the cylinders, where the flow is obviously
affected by the turbulence, it is observed
that $v$ is nearly independent of the
axial coordinate. This is 
confirmed in figure~\ref{profiles_wide} $(b)$,
where $v(r)$ is shown at three different
axial locations. Although these profiles collapse together, they differ substantially
from the desired quasi-Keplerian velocity
profile~\eqref{Couetteflow}, shown as
a (black) solid line in figure~\ref{profiles_wide} $(b)$. Interestingly, despite
vanishing fluctuations (and hence Reynolds stresses) in the bulk, the profile is far
from ideal because of the effect of the global EC.

\subsection{HTX}\label{sec:high_HTX}

Figure~\ref{meridional_high} $(b)$  shows the structure of the time-averaged secondary meridional flow for the \emph{HTX} configuration at $R_s=22529$.
As $R_s$ increases both radial and axial velocities are progressively confined to the end plates,
resulting in an increasingly larger region where the flow remains nearly purely azimuthal.
There is a fraction of the secondary flow generated at the end plates that                          
is transported to the equator over                                                      
a boundary layer at the inner cylinder. The formation of this boundary layer leads                                                
to the emergence of two large-scale circulation cells,                                     
which are progressively confined to the inner cylinder as                               
$R_s$ is increased. These cells are similar to but significantly weaker                               
than those in the \emph{WR} configuration.                     
Turbulent fluctuations                 
are also progressively confined to the end plates as $R_s$ increases (see~figure~\ref{turb_evol} $(b)$).     
Some fluctuations caused by the                                                
large-scale recirculation flow may also occur                                  
in the vicinity of the inner cylinder. However,                                           
given their low intensity,  it can be stated that the flow
remains nearly laminar when sufficiently far from the end plates.

\begin{figure}\setlength{\piclen}{0.16\linewidth}                                                                            
  \begin{center}                                      
    \begin{tabular}{cccc}                                                                   
      $(a)$ & $(b)$ & $(c)$ & $(d)$ \\                                                                   
      \includegraphics[width=\piclen]{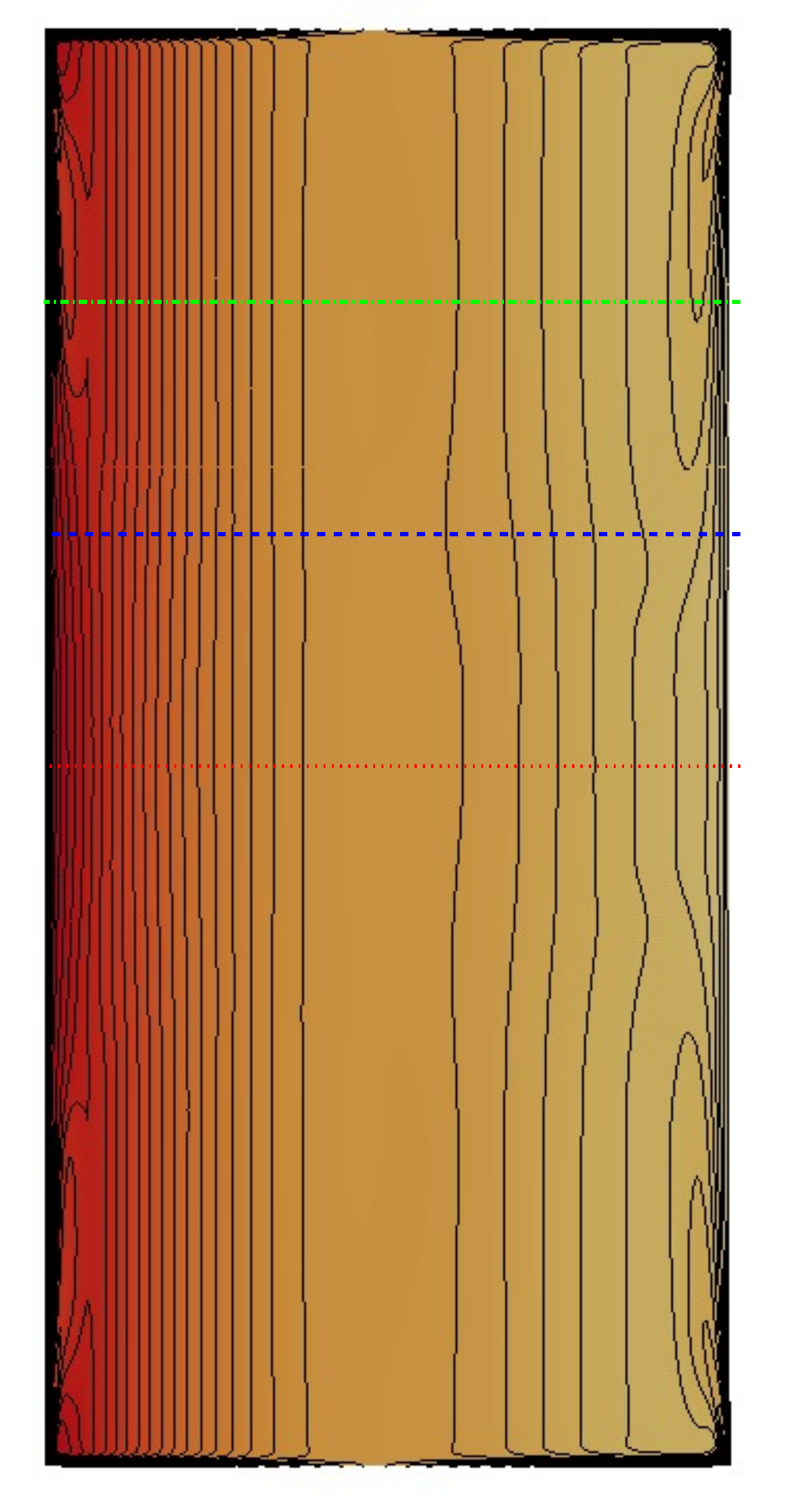} &                           
      \includegraphics[scale=0.35]{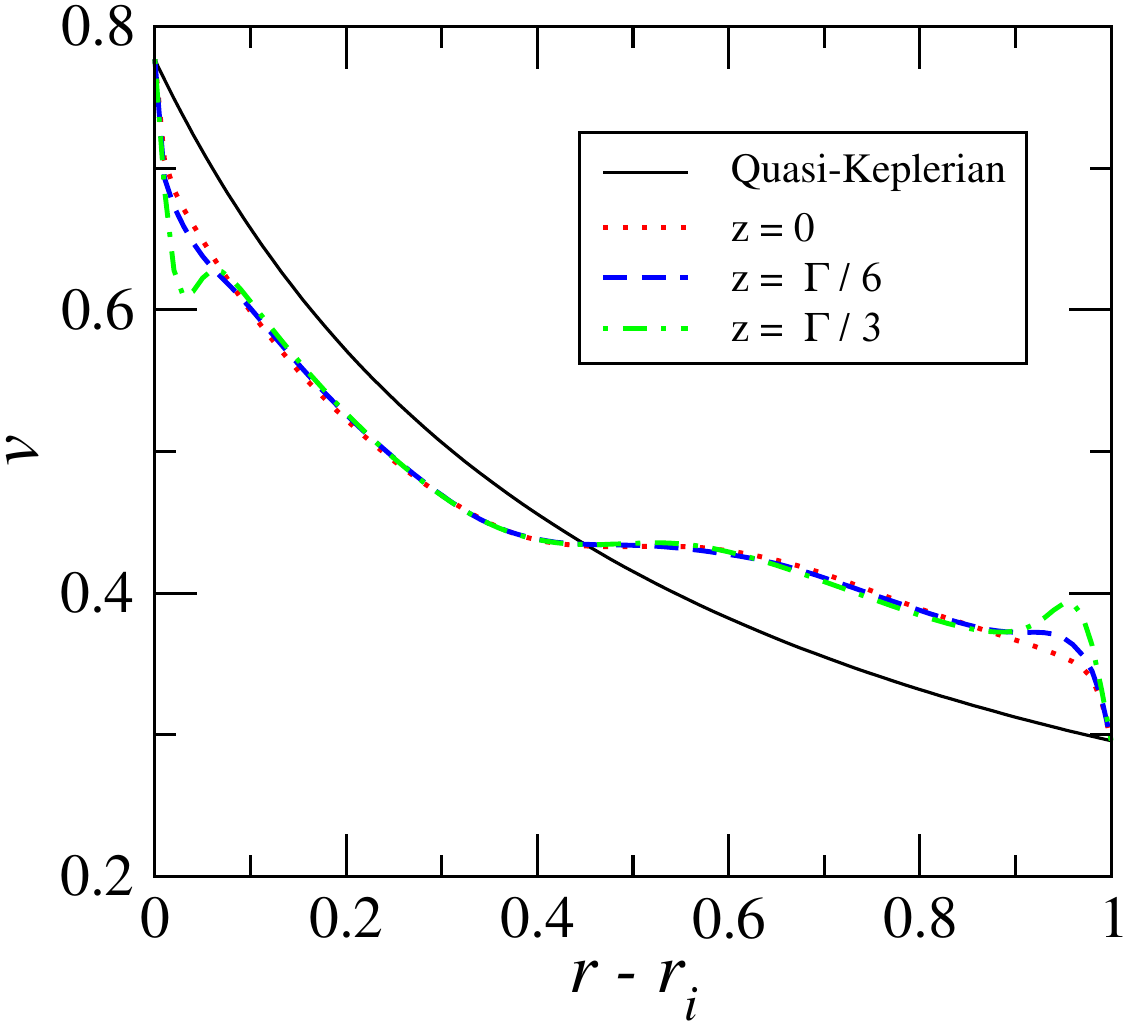}&                                                         
      \includegraphics[width=\piclen]{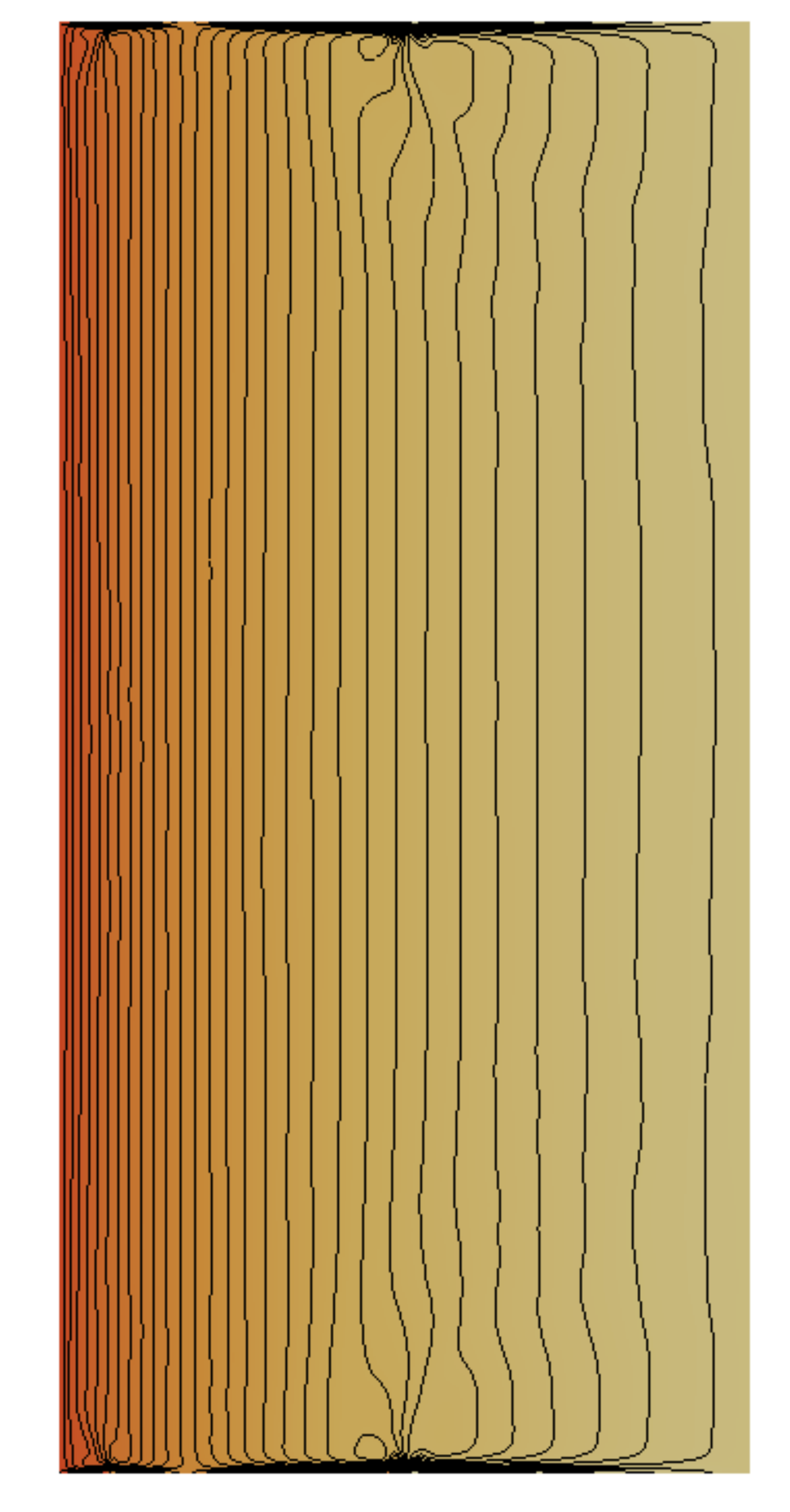} &                           
      \includegraphics[scale=0.35]{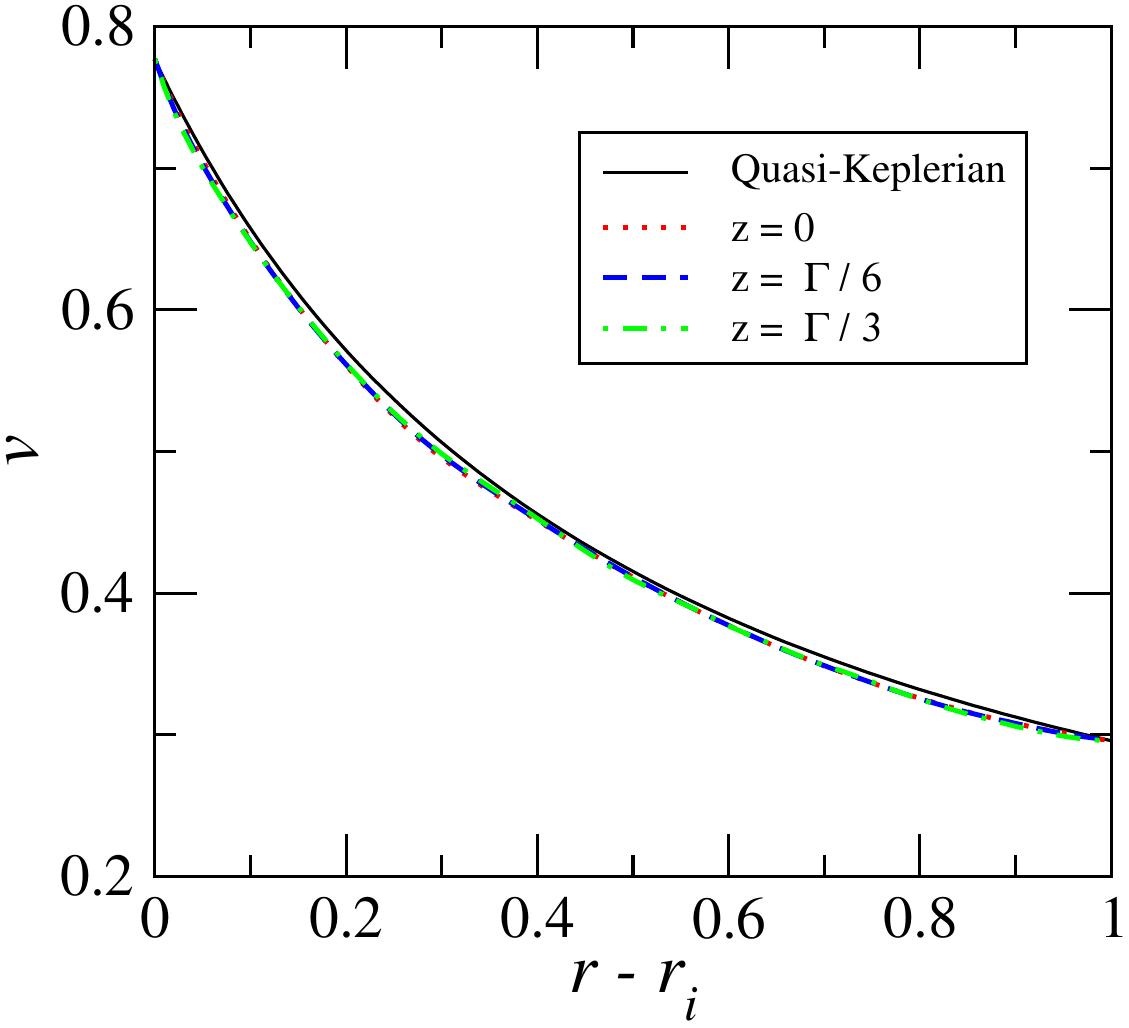}\\
    \end{tabular}                                                                         
  \end{center}                                                                            
  \caption{ $(a)$ and $(c)$ show color maps and contours illustrating the mean azimuthal velocity $v$
    in the \emph{WR} and \emph{HTX} configurations respectively, whereas           
    radial profiles of $v$ computed at three different axial locations are shown          
    in $(b)$ \emph{WR} and $(d)$  \emph{HTX}. In both cases the solution depicted 
    corresponds to the maximum $R_s$ achieved in our simulations, $R_s=47630$ in $(a)$ and $(b)$, and $R_s=32180$ in $(c)$ and $(d)$.
    There are $40$ contours equally distributed across the full range of $v$. The (black) 
    shows the quasi-Keplerian velocity profile~\eqref{Couetteflow} that would be achieved 
    The horizontal lines superimposed upon figure $(a)$ indicate the axial locations at which the azimuthal velocity profiles
    have been computed.}                                                                  
  \label{profiles_wide}                                                                   
\end{figure}

 As seen in figure~\ref{profiles_wide} $(c)$, the mean azimuthal                           
velocity is only significantly affected by the secondary flows                            
in the vicinity of the end plates. Hence, negligible differences                          
were found when measuring radial profiles of the mean azimuthal velocity                  
in the bulk at different axial locations (see figure~\ref{profiles_wide} $(c)$).          
Despite large-scale secondary flows develop at the inner cylinder,                        
unlike for the \emph{WR} configuration,                                            
these are not sufficiently strong as to significantly modify                              
the azimuthal velocity. As a result,                                                      
the azimuthal velocity profiles in this configuration                                     
closely approximate the desired quasi-Keplerian
Couette flow~\eqref{Couetteflow}.
It should be noted that \cite{EdJi14}
reported velocity profiles with a small but noticiable
deviations from~\eqref{Couetteflow} near the inner cylinder.
Such deviations suggest that, while the turbulence at the inner cylinder
does not affect the bulk flow in
our simulations, its influence might not be entirely negligible
in experiments at larger $R_s$.

\subsection{Dynamics of the \emph{MRI} configuration}\label{sec:MRI}

\begin{figure}\setlength{\piclen}{0.16\linewidth}
  \begin{center}
    \begin{tabular}{cccc}
      $(a)$  & $(b)$ & $(c)$ & $(d)$\\
      \includegraphics[width=\piclen]{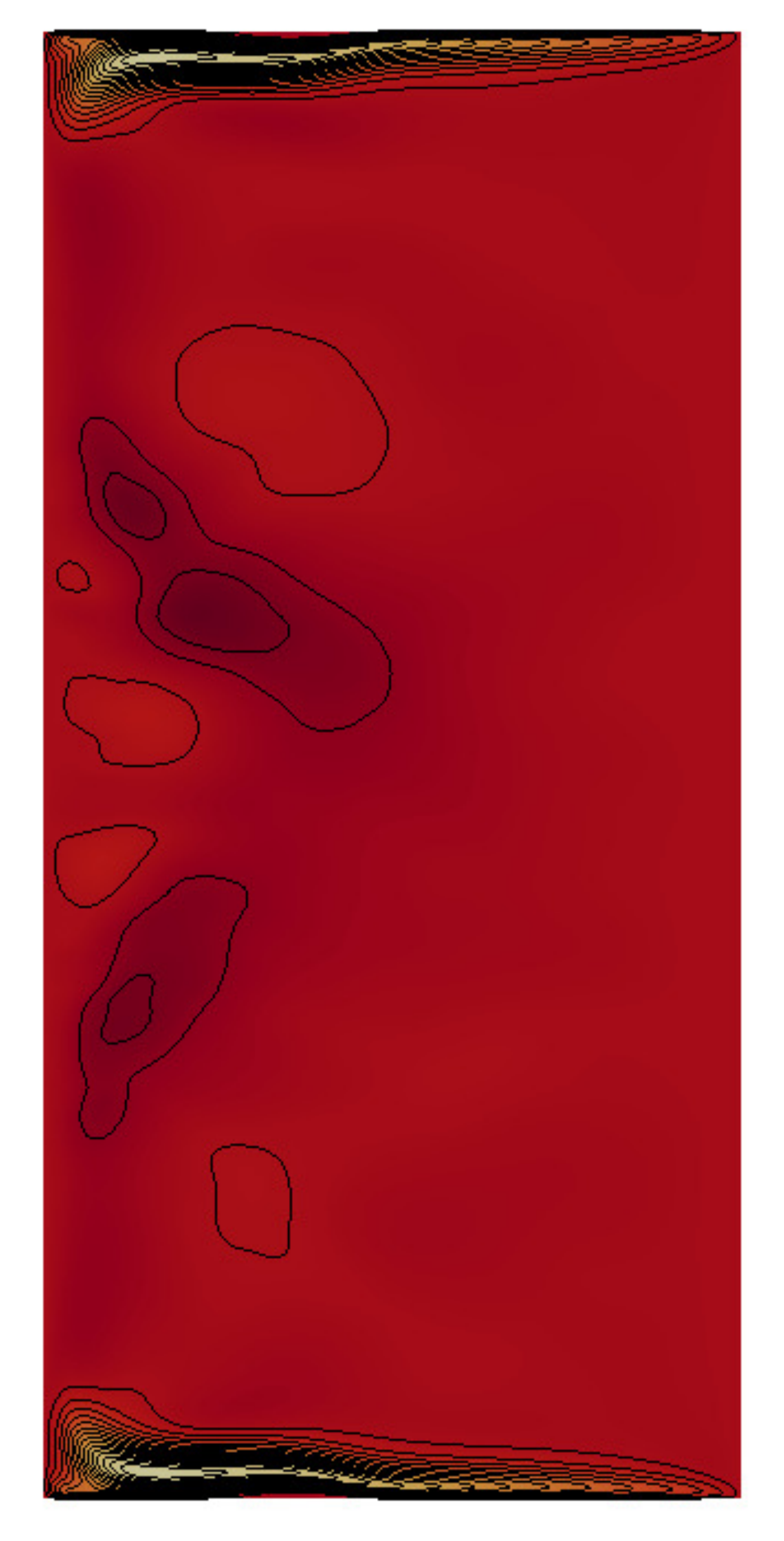} &
      \includegraphics[width=\piclen]{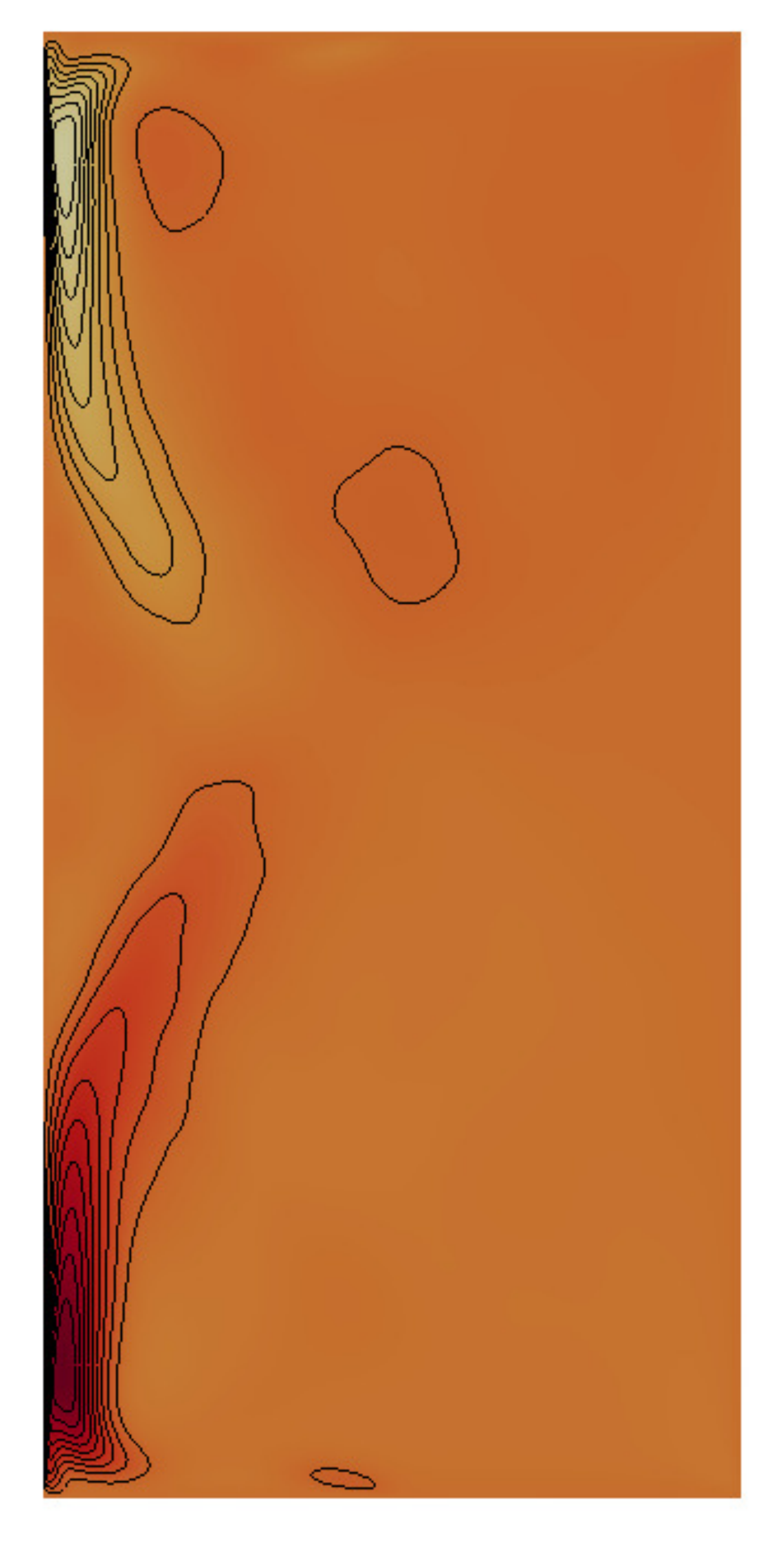}&
      \includegraphics[width=\piclen]{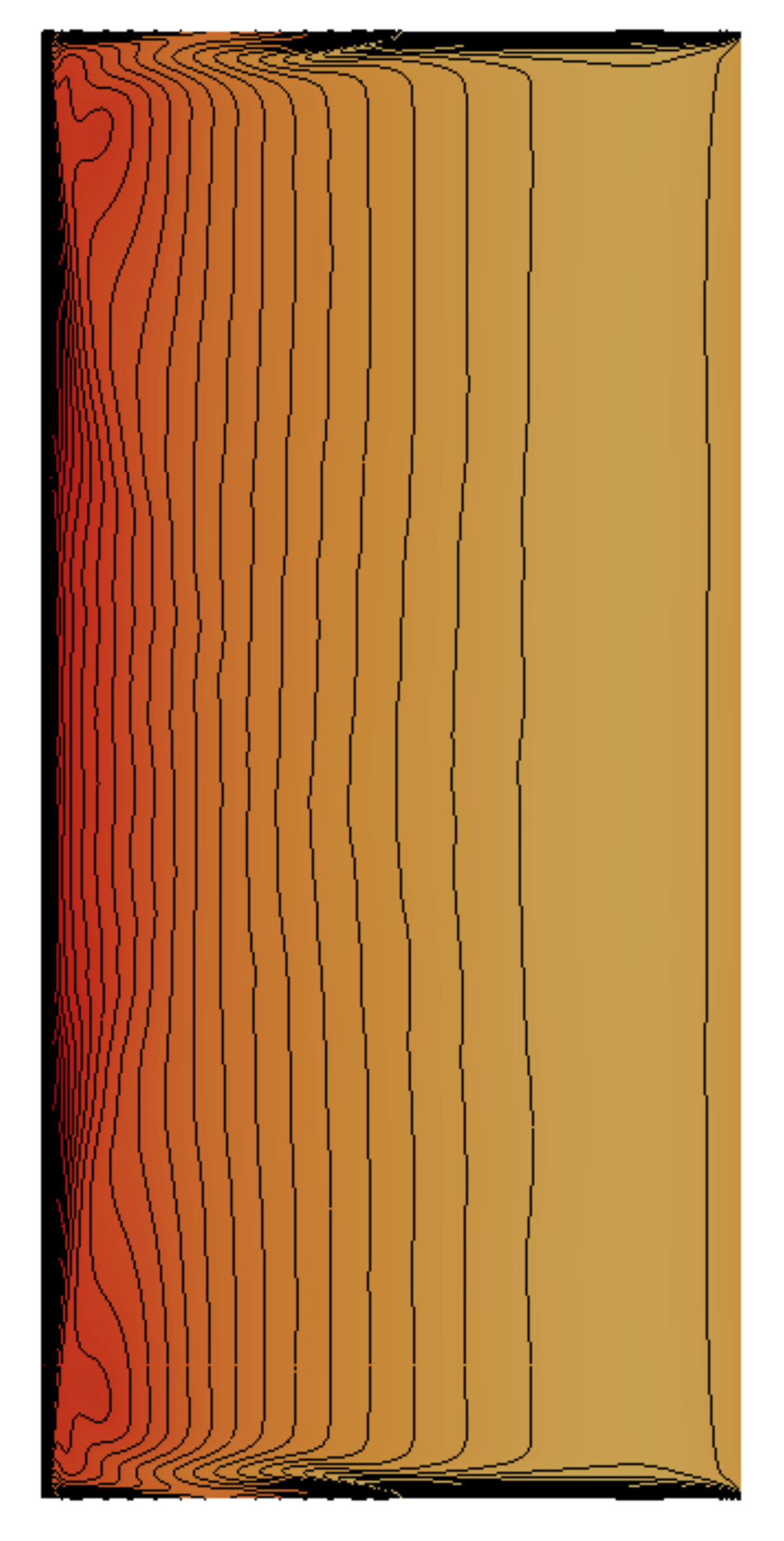} &
      \includegraphics[scale=0.36]{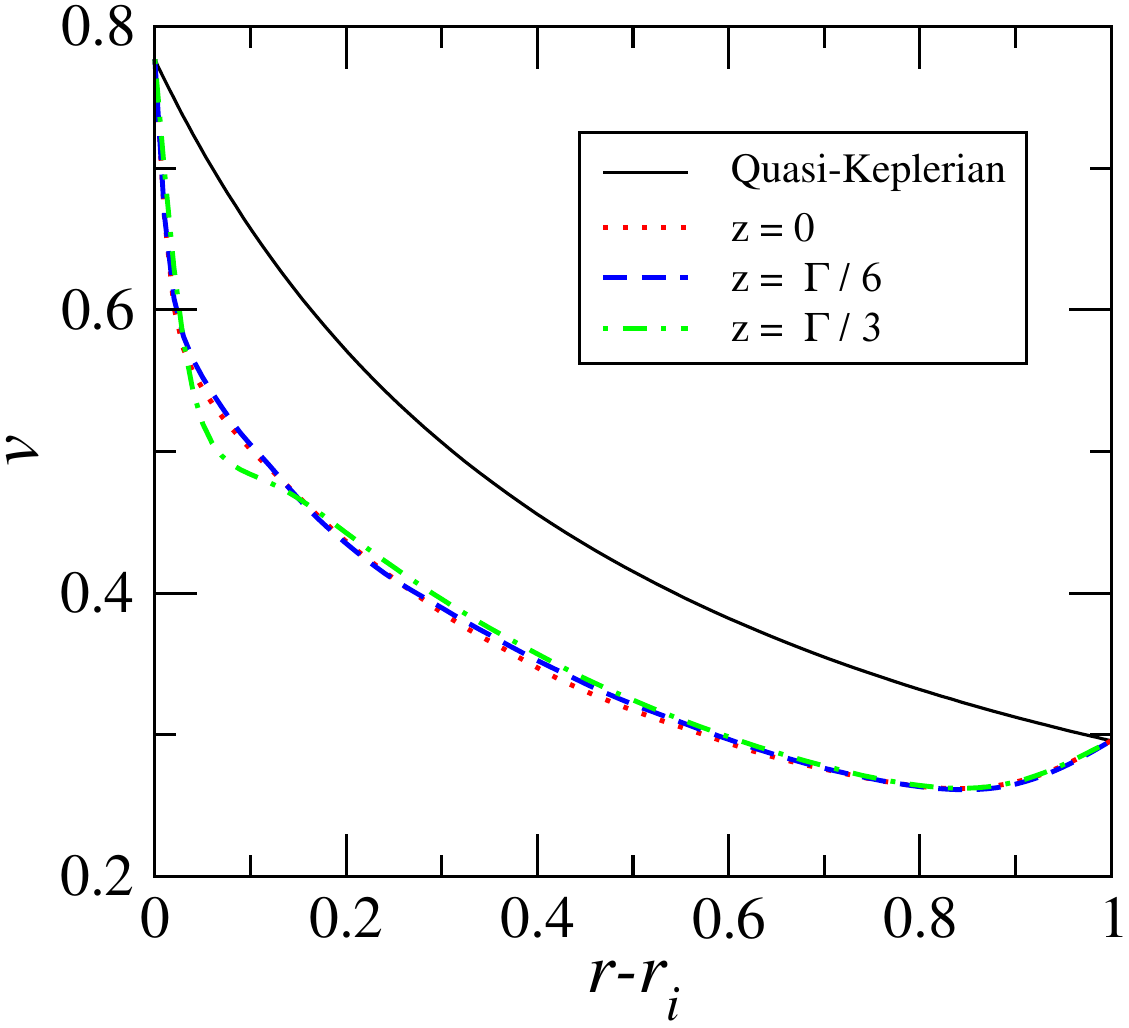}\\
      $u$ & $w$ & $v$ ~ \\
    \end{tabular}
  \end{center}
  \caption{$(a)$, $(b)$ and $(c)$ show meridional sections $(r,z)$ illustrating the mean radial,
    axial and azimuthal velocities in the \emph{MRI} configuration
    for a solution computed at $R_s = 12874$. $20$ contours
    equally distributed across  $u \in [-0.15,0.03]$, $w \in [-0.17,0.17]$ and $v \in [0.06,0.78]$ have been added.
    $(d)$  Radial profiles of $v$ computed at the same axial locations as
    in figure~\ref{profiles_wide}.}
  \label{second_MRI}
\end{figure}

\cite{Av12} performed direct numerical simulations of the \emph{MRI} configuration of \citet{JiBuSchGo06} up to $R_s=6437$,
for which turbulence was found to fill the entire domain.
Here we extend the range of Reynolds numbers by a factor of two.
First of all, we note that the rotation speeds of the end-plate rings used by \cite{JiBuSchGo06} are not optimal
and result in large torque at the end plates. At low $R_s$, the meridional circulation resembles that of the \emph{WR} configuration (see figure 1 in~\cite{Av12}),
with a strong radial jet located at the equatorial region. However, unlike the \emph{WR}, the radial flow at the end plates is entirely inward,
and so there is a single large-scale circulation cell in the upper and lower half of the experiment 
As in the \emph{WR} configuration the large scale circulation cells and
turbulent motions cluster here progressively near the inner cylinder (see figures~\ref{second_MRI} $(a)$ and $(b)$),
leaving a nearly azimuthal and laminar flow in the remaining part of the gap. 
 
 Figures~\ref{second_MRI} $(c)$ and $(d)$ show the time-averaged azimuthal velocity $v$ and radial profiles of $v$ at $R_s = 12874$ respectively.
Here $v(r)$ is also nearly uniform in the axial direction, but differs from the theoretical Couette flow~\eqref{Couetteflow}.
Hence we conclude that although \cite{JiBuSchGo06} measured negligible Reynolds stresses in the bulk of their experiment,
their flows were strongly turbulent in thin boundary layers at the cylinders.
Interestingly, in the bulk the profiles are in fact quasi-Keplerian yet substantially shifted with respect
to Couette flow because of the sharp drop at the cylinder boundary layers.  Similar velocity profiles for this configuration
  were reported in the numerical simulations of ~\cite{obabko2008} and experiments of~\cite{ScJiBuGo12}, who speculated that the deviation in the profiles near the inner cylinder might
    be caused by the existence of turbulent Stewartson boundary layers.

\section{Discussion and conclusions}                                                                     

We have performed direct numerical simulations of the flow in a Taylor-Couette device that \cite{EdJi14}
specifically designed  to infer the hydrodynamic stability of constant-density Keplerian flows.
A first interesting observation is that the occurrence of turbulence at low $R_s$ appears to be
a robust feature of quasi-Keplerian Taylor--Couette flows.
Nevertheless, turbulence manifests itself differently depending on axial boundary conditions.
In the \emph{WR} configuration, as well as in the \emph{HTX} and \emph{MRI} configurations if the boundary conditions are not optimal,
as in \cite{JiBuSchGo06}, the end plates drive a large-scale EC which gives rise to strongly turbulent boundary layers at the cylinders.
In contrast, when the \emph{HTX} configuration is operated under optimal boundary conditions,
the EC and associated turbulence is localised near the end plates. 

As $R_s$ increases turbulence localises to thin boundary layers, whereas the flow in the bulk becomes nearly azimuthal and axially uniform.
The progressive relaminarisation of the bulk flow  observed in these configurations does not however imply that they are all adequate
to infer the stability of astrophysical flows. As reported in~\cite{EdJi15}, the azimuthal velocity profiles achieved in the \emph{WR} configuration
differ substantially from a quasi-Keplerian profile, even in their optimal regime of operation, whereas laminar Couette profiles can be realized in
the \emph{HTX} configuration with optimal boundary conditions.  \cite{LePaAuDaKe16} showed that quasi-Keplerian
  profiles can also be achieved in experiments if stable stratification is added near the end plates.
  However, this method becomes impractical for the large Reynolds numbers investigated in
  the Princeton experiments.

It is remarkable that in spite of the nearly one order of magnitude gap in Reynolds numbers between our simulations
and the experiments of \cite{EdJi15}, the azimuthal velocity profiles are indistinguishable.
This suggests that, despite the turbulent boundary layers, the velocity profiles exhibit self-similar behaviour,
in agreement with the observations~\cite{EdJi15}.
Taken together, these results show that isothermal constant density quasi-Keplerian Taylor--Couette flows are  stable at least up to
$R_s=\mathcal{O}(10^6)$. Noteworthy, for $\eta=0.3478$ and $Re=10^6$  the energy of disturbances imposed to the laminar flow can be
transiently amplified up to a factor of $G=408$ \citep[see eq.~(5.1) in][]{Maretzke_jfm2014}.
Hence it seems that for quasi-Keplerian flows linear transient growth is a poor indicator of turbulence transition. 
                                    
Our results highlight that experiments of astrophysical flows cannot only rely on measuring velocity fluctuations  alone,
because vanishing Reynolds stresses do not imply quasi-Keplerian velocity profiles.
Similarly, torque measurements are inadequate because they cannot be used to infer the level of turbulence in the bulk,
which may be laminar despite highly turbulent boundary layers.

 We remark that the experiments simulated here were performed with a constant density fluid,
 whereas in accretion disks the gas is strongly stratified in the axial and radial directions.
 Hence the results shown here apply to barotropic gases only.
 For stratified quasi-Keplerian flows several instabilities have been reported.
 Examples are the strato-rotational instability~\citep{MoMcWiJaYa01,BaGal07},
 the radiative instability~\citep{LeDiRie10,riedinger2011radiative},
 the zombie vortex instability~\citep{MaPeJiBaLe15,LeLa16},
 the vertical shear instability~\citep{UrBra98,NeGrUm13} and
 the subcritical global instability~\citep{KlaBo03,JoGa06,PeStJu07}.
 The ability to realize quasi-Keplerian velocity profiles
 in experiments at large Reynolds numbers
 opens up avenues for new experimental investigations of some of these instabilities
 and their underlying mechanisms. Candidates
 are the vertical shear instability or
 subcritical global instability, which could be addressed with experiments of radially stratified
 Taylor--Couette flows including thermal relaxation, or the zombie vortex instability,
 which could be studied in a Taylor--Couette setup subject to strong stable stratification in the vertical
 direction. Nevertheless, additional issues related to the interplay between
 end plates and stratification must be first addressed~\citep{LoMaAv15}.

\acknowledgments

We are grateful to red Espa\~nola de Supercomputaci\'on (RES) and the Regionales Rechenzentrum Erlangen (RRZE) for the computational resources provided.

\bibliography{local}
\bibliographystyle{jfm}

\end{document}